\documentclass[journal]{IEEEtran}

\usepackage{amsmath,amssymb}
\usepackage[final]{graphicx}
\usepackage{cite}
\usepackage{subfigure}
\usepackage{comment,color}
\usepackage{subfigmat}

\newcommand{\R}{\ensuremath{{\mathbb R}}}
\newcommand{\Z}{\ensuremath{{\mathbb Z}}}
\newcommand{\N}{\ensuremath{{\mathbb N}}}
\newcommand{\C}{\ensuremath{{\mathbb C}}}
\newcommand{\CC}{{\mathcal C}}
\newcommand{\ra}{\rightarrow}

\newcommand{\uuu}{\mathsf{u}}
\newcommand{\zzz}{\mathsf{z}}
\newcommand{\ttt}{\mathsf{t}}
\newcommand{\nnn}{\mathsf{n}}
\newcommand{\mmm}{\mathsf{m}}
\newcommand{\ppp}{\mathsf{p}}
\newcommand{\qqq}{\mathsf{q}}

\newcommand{\rrr}{\mathsf{r}}
\newcommand{\sss}{\mathsf{s}}
\newcommand{\LLL}{\mathsf{L}}
\newcommand{\RRR}{\mathsf{R}}
\newcommand{\TTT}{\mathsf{T}}
\newcommand{\QQQ}{\mathsf{Q}}

\newcommand{\FF}{\mathbf{F}}
\newcommand{\GG}{\mathbf{G}}
\newcommand{\HH}{\mathbf{H}}
\newcommand{\JJ}{\mathbf{J}}
\newcommand{\PP}{\mathbf{P}}
\newcommand{\QQ}{\mathbf{Q}}
\newcommand{\RR}{\mathbf{R}}

\newcommand{\xx}{\mathbf{x}}
\newcommand{\zz}{\mathbf{z}}
\newcommand{\uu}{\mathbf{u}}
\newcommand{\yy}{\mathbf{y}}
\newcommand{\rr}{\mathbf{r}}
\newcommand{\cc}{\mathbf{c}}

\newcommand{\ol}{\overline}

\newcommand{\col}{\mathrm{col}}

\newcommand{\Enc}{\mathsf{Enc}}
\newcommand{\Dec}{\mathsf{Dec}}
\newcommand{\Add}{\mathsf{Add}}
\newcommand{\Mult}{\mathsf{Mult}}
\newcommand{\IntMult}{\mathsf{IntMult}}

\newcommand{\sk}{\mathsf{sk}}

\newtheorem{thm1}{\bf Theorem}
\newtheorem{prop1}{\bf Proposition}
\newtheorem{lem1}{\bf Lemma}
\newtheorem{asm1}{\bf Assumption}
\newtheorem{defn1}{\bf Definition}
\newtheorem{rem1}{\bf Remark}
\newtheorem{cor1}{\bf Corollary}

\newtheorem{proof1}{\it Proof}

\newenvironment{asm}{\begin{asm1}}{\hfill$\square$\end{asm1}}
\newenvironment{rem}{\begin{rem1}}{\hfill$\square$\end{rem1}}
\newenvironment{lem}{\begin{lem1}}{\hfill$\square$\end{lem1}}
\newenvironment{thm}{\begin{thm1}}{\hfill$\square$\end{thm1}}
\newenvironment{cor}{\begin{cor1}}{\hfill$\square$\end{cor1}}
\newenvironment{prop}{\begin{prop1}}{\hfill$\square$\end{prop1}}
\begin{document}
%
\title{
	\huge
	Dynamic Controller that Operates
	over Homomorphically Encrypted Data
	for Infinite Time Horizon
}


\author{Junsoo Kim, {\it Member, IEEE},
	Hyungbo Shim, {\it Senior Member, IEEE},
	and
	Kyoohyung Han \vspace{-7mm}
\thanks{
This work was supported
by
the National Research Foundation of Korea(NRF) grant funded by the Korea government(Ministry of Science and ICT) (No.~NRF-2017R1E1A1A03070342).
}
\thanks{
	J.~Kim is with the Division of Decision and Control System, KTH Royal Institute of Technology, Sweden.
	H.~Shim is with ASRI, Department of Electrical and Computer Engineering, Seoul National University, Korea.
	K.~Han is with Samsung SDS, Korea. This work is done before K.~Han joins Samsung SDS.
	}
}


\maketitle

\begin{abstract}
In this paper, we present a dynamic feedback controller that computes the next state and the control signal over encrypted data using homomorphic properties of cryptosystems, whose performance is equivalent to the linear dynamic controllers over real-valued data. Assuming that the input as well as the output of the plant is encrypted and transmitted back to the controller, it is shown that the state matrix of any linear time-invariant controller can be always converted to a matrix of integer components. This allows the dynamic feedback controller to operate for infinite time horizon without decryption or reset of its internal state. For implementation in practice, we illustrate the use of a cryptosystem that is based on the Learning With Errors problem, which allows both multiplication and addition over encrypted data. It is also shown that the effect of injected random numbers during encryption for security can be maintained within a small bound by way of the closed-loop stability.
\end{abstract}


%
\IEEEpeerreviewmaketitle

\section{Introduction}\label{sec:intro}

Recent years have seen the development of networked control systems,
and the threat of cyber-attacks has been a major problem.
Against the intrusion of unauthorized access to computing devices in the network,
the use of homomorphic encryption
has been introduced \cite{Kogiso15CDC,Farokhi17CEP,Kim16NECSYS},
so as to protect all the data in the network by encryption
while
the control operation is directly performed on the encrypted data without decryption.
As
the significance of encrypted control
is elimination of the decryption key from the network
so that it decreases the vulnerability,
it has been applied to various targets such as quadratic optimization \cite{Shoukry16CDC}, average consensus \cite{Hadjicostis18CDC},
cooperative control \cite{Darup19CSL},
model predictive control \cite{Alexandru18}, and reset control \cite{Murguia18Arxiv}.

However,
{\it dynamic control operation over encrypted data} has been a challenge,
in which the state of the controller is recursively updated as encrypted variables, and managed without the decryption.
As it requires recursive multiplication with fractional numbers in general,
it is necessary for digital controllers to truncate the significand of the state
from time to time,
to avoid the overflow (see Section~\ref{subsec:recursive}).
Such truncation would be performed by right shift or division of numbers,
but it is impossible
for general homomorphic cryptosystems
to perform the division of messages
an infinite number of times.

As a consequence,
it has been a common concern in the community,
and
several attempts have been made
to deal with
linear dynamic controllers.
For example,
the use of bootstrapping technique \cite{Gentry09} of fully homomorphic encryption is considered in \cite{Kim16NECSYS},
in order to refresh the state for unlimited operation.
However, its computational complexity hinders it from being used in practice.
In \cite{Murguia18Arxiv},
the idea of intermittent reset
of the state to the initial value is proposed, but
it results in degradation of the performance.

In this paper,
we propose a method to perform the operation of linear dynamic controllers over encrypted data,
for infinite time horizon without the reset or decryption of the state.
Following the observation  in \cite{Cheon18} that linear systems having {the state matrix}
of integer components
avoid the overflow problem without the truncation of the significand of the state,
the proposed scheme is based on the conversion of
the state matrix of real numbers into that of integer components
without use of scaling.
	Based on a novel pole-placement technique, all the eigenvalues of the state matrix are appropriately placed
	so that the whole matrix can be transformed to
	have integer components,
	while the system can still have the same input-output
	relation
	compared with the given controller.
By doing so,
given any linear dynamic system,
its operation can be performed over encrypted data for an infinite time horizon,
where the parameters
for quantization and encryption
can be chosen to prevent performance degradation.

For the operation of infinite time horizon,
the proposed method
makes use of re-encryption for the controller output
(rather than re-encrypting the controller state).
	It is based on the rationale that
	the transmission of the controller output to the actuator stage and its decryption at the actuator are necessary for the feedback control, so that it can be re-encrypted at the actuator stage and transmitted back to the controller, assuming that the communication line between the controller and the actuator is bi-directional.
	Since
	the study of encrypted control aims for eliminating or minimizing the use of decryption for control operation,
	the significance of the proposed method is reduction of the amount of data for decryption,
	compared with the existing model as in \cite{Kogiso15CDC} that admits re-encryption of the controller state.	
This will also reduce
communication burden
for additionally transmitting the encrypted state,
and
computational burden (at the actuator)
for discarding the least significant digits during re-encryption.

Another contribution of this work is that
a criterion for choosing the size of underlying space for encrypted data is provided,
which is less conservative than the way used in \cite{Farokhi17CEP}, \cite{Darup19CSL}, or \cite{Cheon18}.
We propose that
selecting the parameter only to cover the range of the controller output is enough to maintain the performance,
despite that some portion of the controller state or input may be lost during the operation (see Remark~\ref{rem:cut}).

The proposed method is applicable with any
cryptosystems that allow additions over encrypted data, such as the Paillier encryption \cite{Paillier},
but the use of recent cryptosystems based on Learning With Errors (LWE) problem \cite{Reg05,GSW13,GSW-LWE} is also suggested.
Compared with the Paillier cryptosystem that has been commonly used for control operations, as in \cite{Farokhi17CEP,Shoukry16CDC,Hadjicostis18CDC,Darup19CSL,Alexandru18,Murguia18Arxiv},
advantages of using LWE-based schemes are listed as follows:
\begin{itemize}
	\item LWE-based schemes have both abilities of addition and multiplication over encrypted data,
	so that it can protect both control parameters and signals, by encryption.

	\item Based on the worst-case lattice problem
	instead of the factoring problem,
	they are known as post-quantum cryptosystems,
	i.e., secure against quantum computers \cite{quantum}.
	\item
	For the case of the Paillier encryption,
	the operation units over encrypted data consist of exponentiations and divisions,
	so that
	it may require computation methods such as Montgomery algorithms in practice, as in \cite{Tran19}.
	In contrast,
	in the case of the LWE-schemes 
	in \cite{GSW13} and \cite{GSW-LWE}
	which utilizes a distinct encryption method for the
	control parameters,
	the units consist of matrix multiplications and bit operations so that it has benefits of simple implementation.
\end{itemize}

To include the case of LWE-based cryptosystems, in this paper,
we use a general additively homomorphic cryptosystem,
in which a random error can be injected to lower bits of a (scaled) message during encryption.
We also show how to use the ability of multiplication over encrypted data,
where the abstraction accords with the cases of using the cryptosystems in \cite{GSW13} and \cite{GSW-LWE}.
The descriptions of the schemes \cite{GSW13} and \cite{GSW-LWE} are also found in Section~\ref{subsec:homomorphic} as examples,
but for more explanations on using them for control operation
with illustrative examples, we refer the readers to \cite{Kim19}.

Although the injection of errors
is a key to the security of the LWE-based encryptions,
in fact,
the errors may grow unbounded
under the recursive operation
and this is another obstacle for the operation of infinite time horizon.
In this paper,
we show
that
the injected errors and their growth
can be regarded as external disturbances or perturbations so that it can be controlled under the closed-loop stability. As a result, it will be seen that the proposed method based on 
LWE-based encryptions
can perform the recursive operation for an infinite time horizon and have the same level of performance.

The organization of the rest of this paper is as follows.
Section~\ref{sec:pre} begins with preliminaries and the problem formulation.
Section~\ref{sec:dynamic} presents the main result
on encrypting dynamic controllers to operate for an infinite time horizon.
Finally, Section~\ref{sec:simulation} illustrates simulation results, and Section~\ref{sec:conclu} concludes the paper.

{\it Notation:}
The set of integers, positive integers, non-negative integers,  real numbers, and complex numbers are denoted by $\Z$, $\N$, $\N_0$, $\R$, and $\C$, respectively.
The floor, rounding, and ceiling function are denoted by $\lfloor\cdot\rfloor$, $\lceil\cdot\rfloor$, and $\lceil\cdot\rceil$, respectively.
The set of integers modulo $q\in\N$ is denoted by $\Z_q:=\{0,1,\cdots,q-1\}$,
and for $a\in\Z$,
we define $a\mod q:= a-\lfloor \frac{a}{q}\rfloor q\in\Z_q$.
The functions defined for scalars, such as $\lceil\,\cdot\,\rfloor$ or $(\,\cdot\mod q)$, can also be used for vectors and matrices as component-wise functions.
The zero-mean discrete Gaussian distribution with standard deviation $\sigma>0$ is denoted by $N(0,\sigma)$.
For real numbers, $\left|\cdot\right|$ denotes the absolute value, and for vectors or matrices, $\|\cdot\|$ denotes the (induced) infinity norm.
For a sequence $v_1,\cdots,v_p$ of column vectors or scalars, we define $\col\{v_i\}_{i=1}^{p} :=[v_1;v_2;\cdots;v_p]:= [v_1^\top,\cdots,v_p^\top]^\top$.
For $m\in\N$ and $n\in\N$,
$I_n\in\R^{n\times n}$ and $0_{m\times n}\in\R^{m\times n}$ denote the identity matrix and the zero matrix, respectively.

\section{Preliminaries and Problem Formulation}\label{sec:pre}

\subsection{Homomorphic Encryption}\label{subsec:homomorphic}
We first describe the cryptosystem to be used throughout the paper.
Consider the set $\Z_q=\{0,1,\cdots,q-1\}$ of integers modulo $q\in\N$,
which is closed under modulo addition, subtraction, and multiplication\footnote{
	For example, for $a\in\Z$ and $b\in\Z$,
	we consider
	$a+b\mod q$ for modular addition and $a\cdot b\mod q$ for modular multiplication.
},
as the space of plaintexts (unencrypted data).
And, let
the set $\CC$ be the space of ciphertexts (encrypted data), and let
$\Enc: \Z_q \ra \CC$ and $\Dec:\CC\ra \Z_q$ be the encryption and decryption algorithms\footnote{A secret key (or a public key) should be an argument of the algorithms $\Enc$ and $\Dec$, but it is omitted for simplicity.}, respectively.
As homomorphic encryption schemes allow operations over encrypted data in which the decryption of the outcome matches the outcome of modular arithmetic over the corresponding plaintexts in $\Z_q$,
we suppose that the cryptosystem under consideration is at least {\it additively homomorphic};
with operations defined in $\CC$, it satisfies the following properties\footnote{
	In fact,
	the property H3 can be obtained from H2 of addition.}.

\begin{enumerate}
	\item[H1:]
	There exists $\Delta_\Enc\ge 0$ such that
	for every $m\in\Z_q$,
	it satisfies
$\Dec(\Enc(m))=m+\Delta\mod q$,
	with some $\Delta\in\Z$ such that $\left|\Delta\right|\le\Delta_\Enc$.
\item[H2:] There exists $\Add:\CC\times\CC\ra\CC$
such that
$\Dec(\Add(\cc_1,\!\cc_2))$
$= \Dec(\cc_1)+\Dec(\cc_2)\!\mod \!q$,
for all $\cc_1\in\CC$ and $\cc_2\in\CC$.
\item[H3:] There exists $\IntMult:\Z\times\CC\ra\CC$
such that
$\Dec(\IntMult(k,\cc))= k\cdot\Dec(\cc)\mod q$,
for all $k\in\Z$ and $\cc\in\CC$.
\end{enumerate}

A remark is made about the presence of the error term $\Delta$ in the property H1;
it is to include recent homomorphic encryption schemes based on Learning With Errors (LWE) problem \cite{Reg05}, such as \cite{GSW13} or \cite{Cheon17ASIACRYPT},
which necessarily inject ``errors'' to every message being encrypted,
for the sake of security.
On the other hand, the use of Paillier encryption \cite{Paillier}
(a typical example of additively homomorphic cryptosystems)
still can be considered,
which satisfies H1 with
$\Delta_\Enc=0$,
and satisfies H2 and H3 as well.

\begin{rem}\label{rem:injected_errors}
	The ``injected error'' $\Delta$ in H1 can be eliminated in practice.
	Indeed, if we use $\Enc_\LLL: m \mapsto \Enc(m/\LLL)$ and $\Dec_\LLL: \cc \mapsto \lceil\LLL\cdot\Dec(\cc)\rfloor$ instead of $\Enc$ and $\Dec$,
	where
	$1/\LLL\in\N$ is such that $1/\LLL>2\Delta_\Enc$,
	we obtain $\Dec_\LLL(\Enc_\LLL(m))= \lceil m+\LLL\Delta\rfloor = m $.
	But
	when it comes to H2 and H3,
	the errors may affect the messages,
	especially
	when encrypted data are updated by the operations for unlimited number of times.
\end{rem}

Additively homomorphic encryption schemes allow both addition and multiplication over encrypted data,
but in H3,
one may want to conceal not only the information in $\cc\in\CC$
but also the multiplier $k\in\Z$.
To this end,
we consider
multiplicatively homomorphic property
as well,
in which a separate algorithm may be used for encrypting the multipliers, in general.
For example, the method presented in \cite{GSW13} or \cite{GSW-LWE}
can be employed,
to make use of
the following property.

\begin{enumerate}
	\item[H4:] There exist $\Enc':\Z_q\ra\CC'$,
	$\Mult:\CC'\times\CC\ra\CC$,
	and $\Delta_\Mult \ge 0$
	such that for every
	$k\in\Z_q$ and $\cc\in\CC$,
	it satisfies
	$\Dec(\Mult(\Enc'(k),\cc)) = k\cdot\Dec(\cc)+\Delta\mod q$,
	with some $\Delta\in\Z$ such that $\left|\Delta\right|\le\Delta_\Mult$.
\end{enumerate}

With the properties H2 and H4,
matrix multiplication over encrypted data can be easily performed.
We abuse notation and write
$\cc_1+\cc_2:=\Add(\cc_1,\cc_2)$ and $\cc'\cdot\cc_1:=\Mult(\cc',\cc_1)$ for ciphertexts $\cc_1\in\CC$, $\cc_2\in\CC$, and $\cc'\in\CC'$, as usual,
and use
$\Enc:\Z_q^{n}\ra\CC^{n}$,
$\Enc':\Z_q^{m\times n}\ra\CC'^{m\times n}$,
$\Dec:\CC^{m}\ra\Z_q^{m}$, and
$+: \CC^m\times\CC^m\ra\CC^m$
to apply the algorithms to each component of matrix (or vector).
Then,
multiplication of a vector $[\cc_j]\in\CC^n$ by a matrix $[\cc'_{ij}]\in\CC'^{m\times n}$ is defined as
$${\small\begin{bmatrix}
\cc'_{11}\!\!&\!\!\cdots\!\!&\!\!\cc'_{1n}\\
\vdots\!\!&\!\!\ddots\!\!&\!\!\vdots\\
\cc'_{m1}\!\!&\!\!\cdots\!\!&\!\!\cc'_{mn}\\
\end{bmatrix}\cdot \begin{bmatrix}
\cc_1\\\vdots\\\cc_n
\end{bmatrix}}:=
{\small\begin{bmatrix}
\cc'_{11}\cdot\cc_1\\\vdots\\\cc'_{m1}\cdot\cc_1
\end{bmatrix}}+\cdots+
{\small\begin{bmatrix}
\cc'_{1n}\cdot\cc_n\\\vdots\\\cc'_{mn}\cdot\cc_n
\end{bmatrix}}
$$
so that for every $K\in\Z_q^{m\times n}$ and $\cc\in\CC^n$,
it satisfies
\begin{equation}\label{eq:MatMult}
\Dec(\Enc'(K)\cdot \cc) = K\cdot\Dec(\cc)+\Delta\mod q
\end{equation}
with some $\Delta\in\Z^n$ such that
$
\|\Delta\|\le  n\Delta_\Mult.
$

Utilizing
the additively homomorphic properties H2 and H3 only,
a property of matrix multiplication analogous to \eqref{eq:MatMult} can be obtained,
having
the multiplier
of the matrix multiplication as plaintext.
Define multiplication of a vector $[\cc_j]\in\CC^n$ by a (plaintext) matrix $[k_{ij}]\in \Z^{m\times n}$ as
$${\small\begin{bmatrix}
k_{11}\!\!&\!\!\cdots\!\!&\!\!k_{1n}\\
\vdots\!\!&\!\!\ddots\!\!&\!\!\vdots\\
k_{m1}\!\!&\!\!\cdots\!\!&\!\!k_{mn}\\
\end{bmatrix}\cdot \begin{bmatrix}
\cc_1\\\vdots\\\cc_n
\end{bmatrix}}:=
{\small\begin{bmatrix}
k_{11}\cdot\cc_1\\\vdots\\k_{m1}\cdot\cc_1
\end{bmatrix}}+\cdots+
{\small\begin{bmatrix}
k_{1n}\cdot\cc_n\\\vdots\\k_{mn}\cdot\cc_n
\end{bmatrix}}
$$
where $k_{ij}\,\cdot\,\cc_j:=\IntMult(k_{ij},\cc_j)$.
Then, 
for every $K\in \Z^{m\times n}$ and $\cc\in\CC^n$,
it satisfies
\begin{equation}\label{eq:IntMatMult}
\Dec(K\cdot \cc) = K\cdot\Dec(\cc)\mod q.
\end{equation}

For an example of homomorphic cryptosystem, the rest of this subsection introduces a scheme presented in
\cite{GSW13} and \cite{GSW-LWE},
which satisfies the properties H1--H4.
The encryption, operation, and decryption algorithms are described as follows.
\begin{itemize}
	\item ${\sf Setup}(1^\lambda)$.
	Choose the standard deviation $\sigma>0$,
	the modulus $q= 2^{q_0}$
	with $q_0\in\N$,
	and $\nu_0\in\N$
	and $d\in\N$ so that $\nu:=2^{\nu_0}$ and
	$ \nu^{d-1}<q\le \nu^d$.
	Choose the dimension $n\in\N$, and
	define the ciphertext spaces $\CC$ and $\CC'$,
	as $\CC = \Z_q^{n+1}$ and $\CC'=\Z_q^{(n+1)\times n'}$,
	where
	$n':=d(n+1)$.
	Return ${\it params}=(q,\sigma,n,\nu,d,n')$.
	\item ${\sf KeyGen}({\it params})$. Generate
	the secret key $\sk\in\Z^n$ as a row vector
	with each component sampled from $N(0,\sigma)$.
	Return $\sk$.
	\item ${\sf Enc}(m\in\Z_q,\sk)$.
	Generate a random column vector $a\in\Z_q^n$ and an error $e\in\Z$
	sampled from $N(0,\sigma)$.
	Compute
	$b=-\sk\cdot a + m + e\mod q$.
	Return $[b;a]\in\Z_q^{n+1}$.
	\item $\Dec(\cc\in\Z_q^{n+1},\sk)$.
	Let $\cc=[b;a]$ where $b\in\Z_q$ and $a\in\Z_q^{n}$.
	Return $ b+ \sk\cdot a \mod q\in\Z_q$.
	\item $\Add(\cc_1\in\Z_q^{n+1},\cc_2\in\Z_q^{n+1})$. Return $\cc_1+\cc_2\mod q$.
	\item $\IntMult(k\in\Z,\cc\in\Z_q^{n+1})$.
	Return $k\cdot\cc\mod q$.
	\item $\Enc'(k\in\Z_q,\sk)$.
	Generate a random matrix $A\in\Z_q^{n\times n'}$,
	and a row vector $E\in\Z^{n'}$
	with each component sampled from $N(0,\sigma)$.
	Compute $B=-\sk\cdot A+E$.
	Return
	$k\cdot [I_{n+1},\nu\cdot I_{n+1},\cdots,\nu^{d-1}\cdot I_{n+1}]+ [B^\top,A^\top]^\top
	\,\mathrm{mod}\, q
	\in\Z_q^{(n+1)\times n'}$.
	\item $\Mult(\cc'\in\Z_q^{(n+1)\times n'},\cc\in\Z_q^{n+1})$.
	For
	$i=0,\cdots,d-1$,
	compute $\cc_i= \lfloor \frac{\cc}{\nu^{i}} \rfloor-\lfloor\frac{\cc}{\nu^{i+1}}\rfloor\nu$
	so that $\cc = \sum_{i=0}^{d-1}\cc_i \nu^{i}$.
	Return $\cc'\cdot\col\{\cc_i\}_{i=0}^{d-1}\mod q\in\Z_q^{n+1}$.
	
\end{itemize}

The homomorphic property of the described cryptosystem is stated as follows,
where,
with some $k_0\in\N$ chosen sufficiently large,
we neglect the probability
that $\left|e\right|>k_0\sigma$,
for every integer $e$ sampled from
$N(0,\sigma)$,
and so, assume $|e| \le k_0 \sigma$.

\begin{prop}\label{prop:H}
	The described scheme satisfies the properties H1--H4, with $\Delta_{\Enc}=k_0\sigma$ and $\Delta_{\Mult}= d(n+1)k_0\sigma\cdot \nu$.
\end{prop}

{\it Proof:} H1)
By construction,
for every $m\in\Z_q$, it satisfies
$\Dec(\Enc(m))= m+e\mod q$ with some $\left|e\right|\le k_0\sigma$.
H2)
Note that $\Dec(\cc)=[1,\sk]\cdot \cc \mod q$, $\forall \cc\in\Z_q^{n+1}$.
Then, for $\cc_1\in\Z_q^{n+1}$ and $\cc_2\in\Z_q^{n+1}$,
it is clear that
$\Dec(\cc_1+\cc_2 \mod q) = [1,\sk]\cdot (\cc_1+\cc_2)\mod q = ([1,\sk]\cdot(\cc_1\mod q)+([1,\sk]\cdot\cc_2\mod q))\mod q=\Dec(\cc_1)+\Dec(\cc_2)\mod q$.
H3) Analogously, given $k\in\Z$ and $\cc\in\Z_q^{n+1}$, it follows that $\Dec(k\cdot \cc \mod q) = k\cdot [1,\sk]\cdot \cc\mod q = k\cdot \Dec(\cc)\mod q$.
H4) In the descriptions of $\Enc'$ and $\Mult$,
observe that
$[1,\sk]\cdot[B^\top,A^\top]^\top=E$
and
$[I_{n+1},\cdots,\nu^{d-1}\cdot I_{n+1}]\cdot \col\{\cc_i\}_{i=0}^{d-1}=\cc$.
Hence,
for every
$k\in\Z_q$ and $\cc\in\Z_q^{n+1}$,
it follows that $\Dec(\Mult(\Enc'(k),\cc))= [1,\sk]\cdot \Mult(\Enc'(k),\cc) \mod q =
E\cdot\col\{\cc_i\}_{i=0}^{d-1}+k\cdot[1,\sk]\cdot\cc \mod q
= k\cdot\Dec(\cc)+ E\cdot\col\{\cc_i\}_{i=0}^{d-1}\mod q$,
where $\|E\|\le n'k_0\sigma$ and $ \|\col\{\cc_i\}_{i=0}^{d-1}\|\le \nu$.
It completes the proof.\hfill$\blacksquare$

\begin{rem}\label{rem:GSW}
	One of the benefits of the described scheme is
	that
	each unit of
	the operations
	can be implemented
	with simple
	modular matrix multiplication,
	in which
	the operations such as $(\,\cdot\, \mathrm{mod}\,q)$ or $\lfloor\cdot / \nu\rfloor$	
	can be performed by bit operations,
	thanks to
	the parameters
	$q$ and $\nu$
	chosen as powers of $2$.
\end{rem}

More explanations and example codes can be found in \cite{Kim19}.

\subsection{Problem Formulation}

Consider a continuous-time plant written by
\begin{align}\label{eq:plant}
\begin{split}
\dot x_p(\ttt) &= f(x_p(\ttt),u_p(\ttt)),\\
y_p(\ttt) &= h(x_p(\ttt)),
\end{split}\qquad \ttt\ge 0,
\end{align}
where $x_p(\ttt)\in\R^{\nnn_p}$ is the state,
$u_p(\ttt)\in\R^{\mmm}$ is the input, 
$y_p(\ttt)\in\R^{\ppp}$ is the output of the plant, 
and $\ttt$ is the continuous time index.
To control the plant \eqref{eq:plant},
suppose that a discrete-time linear time-invariant feedback controller has been designed as follows (without $e_x(t)$, $e_u(t)$, $e_0(t)$):
\begin{subequations}\label{eq:controller_given}
\begin{align}
x(t+1) &= Fx(t) + Gy(t)+Pr(t)+e_x(t),\label{eq:controller_given_state}\\
u(t) &= Hx(t)+Jy(t)+Qr(t)+e_u(t),\label{eq:controller_given_output}\\
x(0) &= x_0+e_{0},\qquad
t=0,1,2,...,\notag
\end{align}
\end{subequations}
where
$y(t):=y_p(t\cdot T_s)\in\R^\ppp$, $t=0,1,...,$ is the plant output discretized with the sampling time $T_s>0$,
$x(t)\in\R^\nnn$ is the controller state with the initial value $x_0\in\R^\nnn$,
$r(t)\in\R^\qqq$ is the reference,
and
$u(t)\in\R^\mmm$ is the controller output.
The discrete-time signal $u(t)$ is fed back to the continuous-time plant \eqref{eq:plant} as
$u_p(\ttt)= u(t)$, for $t\cdot T_s\le\ttt< (t+1)\cdot T_s$.
The terms $\{e_x(t),e_u(t),e_0(t)\}$ indicate perturbations, which represent the error between the designed controller and the implemented controller in practice.
They are caused by quantization and encryption, and will be clarified in Section \ref{sec:dynamic}.
Now, let us denote the states and the outputs of \eqref{eq:plant} and \eqref{eq:controller_given} by $x'_p(\ttt)$, $y'_p(\ttt)$, $x'(t)$, and $u'(t)$, when all $e_x(t)$, $e_u(t)$, and $e_0(t)$ are zero.
Here, all the matrices $F$, $G$, $P$, $H$, $J$, and $Q$ are supposed to consist of rational numbers, since any irrational number can be approximated by a rational number with arbitrary precision.

Throughout the paper,
	the closed-loop system is assumed to be (locally) stable.
	First,
	the state, input, and output of \eqref{eq:controller_given} for the ideal case are assumed to be bounded in the closed-loop;
there exists a constant $M>0$, which is known,
such that
\begin{equation}\label{eq:bound}
\|[y'(t);r(t);x'(t);u'(t)] \|\le M,
\end{equation}
for all $t\in\N_0$.
And,
regarding the stability with respect to perturbations,
the following assumption is made.

\begin{asm}\label{asm:stability}
	The closed-loop system is stable,
	with respect to $\{e_x(t),e_u(t),e_0\}$,
	in the sense that
given $\epsilon > 0$,
there exists $\eta(\epsilon) > 0$ such that
if
$\|e_0\|\le \eta(\epsilon)$,
$\|e_x(t)\|\le \eta(\epsilon)$,
and
$\|e_u(t)\|\le \eta(\epsilon)$
for all $t\in\N_0$,
then
\begin{subequations}\label{eq:asm2}
\begin{align}
\|x_p(\ttt)-x'_p(\ttt)\|\le\epsilon,\qquad &\|y_p(\ttt)-y'_p(\ttt)\|\le\epsilon,\label{eq:asm2.1}\\
\|x(t)-x'(t)\|\le\epsilon,\qquad&~\,\|u(t)-u'(t)\|\le\epsilon,\label{eq:asm2.2}
\end{align}
\end{subequations}
	for all $\ttt\ge 0$ and
	$t\in\N_0$,
	respectively.
\end{asm}

The models of the plant and controller are further specified,
in terms of
access to the secret key of cryptosystem, their abilities,
and
constraints on the communication between them:
\begin{itemize}
\item

The plant has access to the encryption and decryption algorithms, with the secret key.
For each sampling time $t$,
the plant encrypts the information of the plant output $y(t)$ and transmit it to the controller.

\item
The controller does not have access to the secret key and decryption, and stores the control parameters (such as the matrices and the initial state of \eqref{eq:controller_given}) and its state, as encrypted.
For each time $t$,
with the received encrypted signal of $y(t)$ as input,
it computes both the next state and the output directly over encrypted data,
only using the homomorphic properties of cryptosystem.
The output of the controller as the operation outcome at time $t$,
which should correspond to encrypted data of the output $u(t)$ of \eqref{eq:controller_given},
is transmitted to the plant and then decrypted for the feedback input of \eqref{eq:plant}.

\item Only the encrypted signal of the controller output $u(t)$ can be transmitted to the plant for decryption.
{Especially, the state of the controller is not allowed to be transmitted and decrypted at the plant, for the whole time}.

\item
Instead,
the plant input $u(t)$ (the decrypted controller output) 
can be re-encrypted and transmitted to the controller,
in addition to the encrypted plant output $y(t)$,
so that it can be
utilized for the update of the controller state.
\end{itemize}

Finally, the problem of this paper is posed.
Given a linear dynamic system \eqref{eq:controller_given},
the objective is to design a dynamic controller
subject to the specified model,
{\it which can perform the operation over encrypted data for an infinite time horizon and has the equivalent performance to \eqref{eq:controller_given} for the whole time};
given $\epsilon>0$,
the decrypted output $u(t)$ of the designed controller should satisfy $\|u(t)- u'(t)\|\le \epsilon$, for all $t\in\N_0$.

\subsection{Controller over Quantized Numbers}
We further specify
the operation of the controller \eqref{eq:controller_given} to digital arithmetic over quantized signals.
Let the inputs $y(t)\in\R^{\ppp}$ and $r(t)\in\R^\qqq$ of \eqref{eq:controller_given} be quantized to integers, as
\begin{equation}\label{eq:input_quantization}
\ol y(t) := \left\lceil\frac{y(t)}{\rrr_1} \right\rfloor\in\Z^\ppp,\qquad \ol r(t) :=  \left\lceil\frac{r(t)}{\rrr_1} \right\rfloor\in\Z^\qqq,
\end{equation}
where $\rrr_1>0$ is the step size for the quantization.
And, since
the parameters of \eqref{eq:controller_given} 
consist of rational numbers,
with some positive integers $1/s_1 \in\N$ and $1/s_2\in\N$, they satisfy

\vspace{-3mm}
{\small
\begin{align*}
\ol F &:= \frac{F}{\sss_1}\in\Z^{\nnn\times\nnn},~ \ol G := \frac{G}{\sss_1}\in\Z^{\nnn\times\ppp},~\ol P:= \frac{P}{\sss_1}\in\Z^{\nnn\times \qqq},\\
\ol H &:= \frac{H}{\sss_2}\in\Z^{\mmm\times\nnn},~ \ol J := \frac{J}{\sss_1\sss_2}\in\Z^{\mmm\times\ppp},~\ol Q:= \frac{Q}{\sss_1\sss_2}\in\Z^{\mmm\times \qqq},
\end{align*}}
and $\ol x_0:= x_0/(\rrr_1\sss_1)\in\Z^\nnn$. Now,
we rewrite
the system \eqref{eq:controller_given} 
as
\begin{subequations}\label{eq:quantized_rev}
\begin{align}\label{eq:quantized_rev_controller}
\begin{split}
\ol x(t+1) &= \lceil \sss_1 \ol F\ol x(t) \rfloor + \ol G \ol y(t) + \ol P \ol r(t),\quad \ol x(0) = \ol x_0
\\
\ol u(t) &= \ol H \ol x(t) + \ol J \ol y(t) + \ol Q \ol r(t),
\end{split}
\end{align}
so that it operates over quantized numbers, as $\ol x(t) \in\Z^\nnn$ and $\ol u(t)\in \Z^\mmm$ for all $t\in\N_0$.

Regarding the feedback of the output $\ol u(t)$ of \eqref{eq:quantized_rev_controller} to the plant,
we consider quantization at the actuator, as well.
First, from the inputs of \eqref{eq:quantized_rev_controller} scaled with the factor $1/\rrr_1$, as \eqref{eq:input_quantization}, and the matrices in \eqref{eq:quantized_rev_controller} with the scale factors $1/\sss_1$ and $1/\sss_2$, it can be seen that the signals $\ol x(t)$ and $\ol u(t)$ are of scale $1/(\rrr_1\sss_1)$ and $1/(\rrr_1\sss_1\sss_2)$, respectively; i.e., $\ol x(t)$ and $\ol u(t)$ have approximate values of $x(t)/(\rrr_1\sss_1)$ and $u(t)/(\rrr_1\sss_1\sss_2)$, respectively.
Thus, the input $u(t)$ of the plant \eqref{eq:plant} can be obtained from $\ol u(t)$, as
\begin{equation}\label{eq:output_quantization}
u(t) = \QQQ^u(\ol u(t)) := \rrr_2\left\lceil\frac{\rrr_1\sss_1\sss_2\cdot\ol u(t)}{\rrr_2}\right\rfloor,
\end{equation}
\end{subequations}
where $\rrr_2>0$ is the step size of quantizer for the plant input.

The following proposition states that the performance of the controller \eqref{eq:quantized_rev} is equivalent to that of \eqref{eq:controller_given}, when the parameters $1/\rrr_1$ and $1/\rrr_2$ for the quantization is chosen sufficiently large.

\begin{prop}\label{prop:quantized_rev}
	Under Assumption \ref{asm:stability},
	there exists a continuous function $\alpha(\rrr_1,\rrr_2) \in \R$
vanishing at the origin
	such that
if	$\alpha(\rrr_1,\rrr_2)\le \eta(\epsilon)$,
then the controller
	\eqref{eq:quantized_rev}
	guarantees that \eqref{eq:asm2} holds,
	for all $\ttt\ge 0$ and $t\in\N_0$.
\end{prop}

{\it Proof:}
We identify the system \eqref{eq:quantized_rev} with the system \eqref{eq:controller_given};
with the perturbations $\{e_x(t),e_u(t), e_0\}$ in \eqref{eq:controller_given} determined as

\vspace{-3mm}
{\small
\begin{align*}
e_x(t) &= {\small\left(\rrr_1\sss_1 \left\lceil\frac{Fx(t)}{\rrr_1\sss_1} \right\rfloor-Fx(t)\right)}+
G{\small\left( \rrr_1 \left\lceil\frac{y(t)}{\rrr_1} \right\rfloor - y(t)\right)} \\&\quad+ P{\small\left(\rrr_1  \left\lceil\frac{r(t)}{\rrr_1} \right\rfloor-r(t)\right)},\\
e_u(t) &= J{\small\left(\rrr_1  \left\lceil\frac{y(t)}{\rrr_1} \right\rfloor -  y(t)\right)} + Q{\small\left(\rrr_1 \left\lceil\frac{r(t)}{\rrr_1} \right\rfloor -  r(t)\right)}\\
&\quad +\left(\rrr_2 \left\lceil\frac{\rrr_1\sss_1\sss_2\ol u(t)}{\rrr_2} \right\rfloor - \rrr_1\sss_1\sss_2 \ol u(t)\right),
\end{align*}}
and $e_0 = 0$, they satisfy
$x(t) =\rrr_1\sss_1 \ol  x(t)$ and
$u(t)= \QQQ^u(\ol u(t))$,
$\forall t\in\N_0$.
Assumption~\ref{asm:stability} completes the proof,
with
\begin{equation}\label{eq:alpha}
\left\| \begin{bmatrix}
e_x(t)\\e_u(t)
\end{bmatrix}\right\| 
\le \left\|\frac{1}{2}\begin{bmatrix}
\rrr_1\sss_1+\|[G, P]\|\rrr_1 \\ \rrr_2 + \| [J, Q]\|\rrr_1
\end{bmatrix}\right\|=:\alpha'(\rrr_1,\rrr_2,\sss_1),
\end{equation}
and $\alpha(\rrr_1,\rrr_2):=\alpha'(\rrr_1,\rrr_2,\sss_1)$.
\hfill$\blacksquare$

\subsection{Recursive Multiplication by Fractional Numbers}\label{subsec:recursive}
Finally, before the main result of this paper is presented,
the difficulty of implementing dynamic controllers over encrypted data and the limitation on recursive multiplication by non-integer real numbers are briefly reviewed.

We revisit the controller \eqref{eq:quantized_rev_controller},
where
recursive multiplication of the controller state, by the matrix $\ol F$, is found.
Since the state matrix $F\in\R^{\nnn\times\nnn}$ would not consist of integers in general,
we may suppose that the scale factor $1/\sss_1\in\N$ for $F$, which makes $\ol F = F/\sss_1\in\Z^{\nnn\times \nnn}$, is not equal to $1$, i.e., $1/\sss_1 > 1$, so that it keeps the fractional part of the matrix $F$.
Then, it follows that the multiplication of $\ol x(t)$ by $\ol F$, which can be seen as multiplication of the significand of $x$ and $F$, respectively, increases the size of the significand of the state; in other words, the scale of $\ol x(t)$, which would be of $1/(\rrr\sss_1)$, will increase to $1/(\rrr\sss_1^2)$, by the multiplication $\ol F \cdot \ol x(t)$.
Thus,
the operation $\lceil \sss_1(\cdot ) \rfloor$ in \eqref{eq:quantized_rev_controller}
divides the outcome $\ol F \ol x(t)$ by $1/\sss_1\in\N$, and truncates the fractional part, so that it keeps the scale of significand for
$\ol x(t+1)$, as the same.

However,
due to
the operation $\lceil \sss_1(\cdot ) \rfloor$,
it is not straightforward to implement \eqref{eq:quantized_rev_controller} directly over encrypted data.
This is because it is generally not possible, for cryptosystems having only the properties H1--H4 of addition and multiplication, to perform the division of numbers over encrypted data.

And especially, when it comes to operation for an infinite time horizon, it follows that dynamic controllers implemented in the usual way, as \eqref{eq:quantized_rev}, are incapable of operating over encrypted data for an infinite time horizon.
One may consider running \eqref{eq:quantized_rev}
without the operation $ \lceil \sss_1(\cdot)\rfloor$, as in \cite{Kim16NECSYS} and \cite{Murguia18Arxiv},
as
$$\ol x_e(t+1) = \ol F \ol x_e(t) + \frac{1}{\sss_1^{t+1}} \left(\ol G\ol y(t)  +  \ol P\ol r(t)\right), $$
where the state $\ol x_e(t)$ has the value of $x(t)/(\rrr\sss_1^{t+1})$, in which the exponent for the factor $1/\sss_1$ increases as time goes by.
But then, the state $\ol x_e(t)$ as an encrypted message cannot be divided by the factor $1/\sss_1$, since only the addition and multiplication is allowed over encrypted data.
Therefore,
the norm of $\ol x_e(t)$ eventually goes to infinity as the time goes by, causing an overflow problem in finite time, and resulting in the incapability of operating for an infinite time horizon.

\begin{rem}
	In fact, it is known that such divisions by scaling factors is possible
	an unlimited number of times,
	when the ``bootstrapping'' techniques of fully homomorphic encryptions \cite{Gentry09} are employed.
	However, as the complexity of the bootstrapping process hinders them from being used for real time feedback control,
	one of the objectives of this paper is not to make use of bootstrapping algorithms.
\end{rem}

\begin{rem}\label{rem:state decryption}
	A direct way to run \eqref{eq:quantized_rev_controller} over encrypted data, for infinite time horizon, is to transmit the updated state as well as the output to the plant side at every time step. This is because the operation $\lceil \sss_1(\cdot) \rfloor$ can be done after decryption as plaintexts.
	However, as it requires additional transmission throughput and computational burden at the plant side, proportional to the dimension of the state,
	we do not admit the state decryption, in this paper.
\end{rem}

So far, we have seen
that encrypting dynamic controllers is not straightforward, and that
the problem is attributable to
recursive multiplication by a non-integer state matrix.
Indeed, 
suppose that all the components of the state matrix $F$ are integers, and the scale factor $1/\sss_1$ is equal to $1$.
Then, the operation \eqref{eq:quantized_rev_controller} is composed of only addition and multiplication over integers,
so that it is possible
to operate over encrypted data, for an infinite time horizon,
only exploiting the additive and multiplicative properties H1--H4 of the cryptosystem.

From this observation,
the approach of this paper is {\it to convert
the state matrix of the given controller to integers,
without use of scaling}.
In the next section,
it will be seen that given any system \eqref{eq:controller_given},
it can be converted to a system having the same input-output relation,
which does not recursively multiply fractional numbers to the state,
so that it can
be encrypted to operate for an infinite time horizon.

\section{Main Result}\label{sec:dynamic}
In this section,
a method for
converting the state matrix of \eqref{eq:controller_given} to integers,
and
a method for
converting the whole system \eqref{eq:controller_given} to a system over the space $\Z_q$
are proposed.
Then,
we show how to utilize the homomorphic properties of the cryptosystem to perform the operation of the converted controller. The performance of the proposed controller is to be analyzed,
and the ability of operating for infinite time horizon is to be seen.

\subsection{Conversion of State Matrix}
Given the controller \eqref{eq:controller_given} where the state matrix $F\in\R^{\nnn\times\nnn}$ consists of rational numbers,
a direct way to keep the significand of $F$, together with its fractional parts, is to use a scale factor $1/\sss_1\in\N$ so that
$\ol F= F/\sss_1\in\Z^{\nnn\times\nnn}$,
as in \eqref{eq:quantized_rev_controller}.
Instead, with the motivation seen in Section \ref{subsec:recursive}, we propose a method for converting $F$ to integers, without scaling.

We first consider converting the state matrix by coordinate a transformation;
even if the given state matrix $F$ does not consist of integers,
there may exist an invertible matrix
$T\in\R^{\nnn\times \nnn}$
such that the state matrix with respect to the transformed state $Tx\in\R^\nnn$ consists of integers, i.e., $TFT^{-1}\in\Z^{\nnn\times\nnn}$.
Considering that the entries of the state matrix determine the eigenvalues of the matrix,
the following proposition is found,
as a tool for the conversion.
\begin{prop}\label{prop:eigenvalue}
	Given $A\in\R^{\nnn\times\nnn}$, there exists $T\in\R^{\nnn\times\nnn}$ such that $TAT^{-1}\in\Z^{\nnn\times\nnn}$, if every eigenvalue
	of $A$ is such that both the real part
	and the imaginary part
	are integers.
\end{prop}

{\it Proof:}
Let
$\lambda_1,\cdots,\lambda_{\nnn-2m},\sigma_1\pm j\omega_1 ,\cdots, \sigma_{m}\pm j\omega_{m}$
be the eigenvalues of $A$,
where
$\lambda_i\in\Z$,
$\sigma_i\in\Z$ and $\omega_i\in\Z$.
It is clear that the transformation of $A$ to the modal canonical form
yields a matrix consisting of integers;
there exists $T\in\R^{\nnn\times\nnn}$ such that
the matrix $TAT^{-1}\in\R^{\nnn\times\nnn}$ takes the form of
{{
	\[
	{\small\begin{bmatrix}
	\begin{matrix}\lambda_1 &\gamma_1&\cdots&0\\
	0&\lambda_2&\ddots&\vdots\\
	\vdots&\ddots&\ddots&\gamma_{\nnn-2m-1}\\
	0&\cdots&0&\lambda_{\nnn-2m}
	\end{matrix}
	& \vrule & 0_{(\nnn-2m)\times 2m} \\
	\hline
	0_{2m\times (\nnn-2m)} & \vrule &
	\begin{matrix}\Lambda_1 &\Gamma_1&\cdots&0_{2\times 2}\\
	0_{2\times 2}&\Lambda_2&\ddots&\vdots\\
	\vdots&\ddots&\ddots&\Gamma_{m-1}\\
	0_{2\times 2}&\cdots&0_{2\times 2}&\Lambda_{m}
	\end{matrix}
	\end{bmatrix}},
	\]
}}
where $\gamma_i=0$ or $1$, 
$$\Lambda_i = {\small\begin{bmatrix}
\sigma_i&\omega_i\\
-\omega_i &\sigma_i
\end{bmatrix}}\in\Z^{2\times 2}, 
\quad \text{and} \quad \Gamma_i = 0_{2 \times 2}~\text{or}~I_{2} \qquad \blacksquare
$$

Proposition~\ref{prop:eigenvalue} can be
considered
as a sufficient condition for transforming the state matrix to integers.
However, in fact,
transformation by itself is not enough to cover all the cases;
the condition $TFT^{-1}\in\Z^{\nnn\times\nnn}$
implies that
the coefficients of
its characteristic polynomial
$\det(sI_\nnn-TFT^{-1})$
are integers,
but due to the invariance $\det(sI_\nnn-F)=\det(sI_\nnn-TFT^{-1})$,
such $T$ does not exist when the coefficients of $\det(sI_\nnn-F)$ are not all integers.

To cover the whole class of linear systems
for the encryption and the operation
for infinite time horizon,
we suggest that
the output $u(t)$ of the controller be treated as an auxiliary input, simultaneously.
Observe that,
from $u(t) = Hx(t) + Jy(t)+Qr(t)$,
the right-hand-side of
\eqref{eq:controller_given_state} can be re-written as
\begin{align}\label{eq:outward}
&Fx(t) + Gy(t) + Pr(t)\\
&= (F\!-\!RH)x(t) + (G\!-\!RJ)y(t)+(P\!-\!RQ)r(t) + Ru(t),\notag
\end{align}
with a matrix $R\in\R^{\nnn\times\mmm}$,
where we suppose $e_x(t)=0$ and $e_u(t)=0$ for simplicity.
Then, by regarding the signal $u(t)$ as an external input of the controller,
the coordinate transformation for converting the state matrix to integers can be found with respect to the new state matrix $F-RH$.

Then,
since the value of the matrix $R$ does not affect the performance of the controller,
 it can be freely chosen for the conversion;
we consider pole-placement design for $R$ so that the matrix $F-RH$ can be transformed to integers, by change of coordinate.
For instance, if
all the eigenvalues of $F-RH$ can be chosen to be integers,
then Proposition~\ref{prop:eigenvalue} will guarantee that there exists $T\in\R^{\nnn\times\nnn}$ such that
$T(F-RH)T^{-1}\in\Z^{\nnn\times\nnn}$, i.e.,
the given controller \eqref{eq:controller_given} can be converted to have the state matrix as integers, without scaling.

As the pole-placement design for the matrix $R$ requires observability of the pair $(F,H)$,
which may not be observable in general,
we consider Kalman observable decomposition for the controller;
with an invertible matrix $W=[W_1^\top,W_2^\top]^\top\in\R^{\nnn\times\nnn}$ such that $z_1 = W_1 x\in\R^{\nnn'}$ and $z_2 = W_2 x\in\R^{\nnn-\nnn'}$ where $\nnn'\le \nnn$,
the controller \eqref{eq:controller_given} is transformed into the form
\begin{subequations}\label{eq:controller_transformed}
	\begin{align}
	z_1(t+1)&= F_{11}z_1(t) + W_1( Gy(t) + Pr(t) + e_x(t))\label{eq:controller_transformed_reduced}\\
	\begin{split}\label{eq:controller_transformed_dispensable}
	z_2(t+1)&= F_{21}z_1(t) + F_{22}z_2(t)\\&\quad + W_2( Gy(t)+ Pr(t) + e_x(t))
	\end{split}
	\\
	u(t) &= H_1 z_1(t) + Jy(t) + Qr(t) + e_u(t),\label{eq:controller_transformed_output}
	\end{align}
\end{subequations}
in which the pair $(F_{11}, H_1)$ is observable.
Then, 
as the sub-state $z_2$ is not reflected in the output $u(t)$ of the controller,
we can remove out the part \eqref{eq:controller_transformed_dispensable} and obtain the ``reduced'' controller \eqref{eq:controller_transformed_reduced} with \eqref{eq:controller_transformed_output}.

Finally,
the observability of $(F_{11},H_1)$ enables the design of $R_1\in\R^{\nnn'\times\mmm}$ such that all the eigenvalues of $F_{11}-R_1H_1$ are integers,
so that
there exists $T_1\in\R^{\nnn'\times\nnn'}$ such that $T_1(F_{11}-R_1H_1)T_1^{-1}\in\Z^{\nnn'\times\nnn'}$,
thanks to Proposition~\ref{prop:eigenvalue}.
As a result,
the following lemma
ensures that
given any
controller \eqref{eq:controller_given},
the state matrix $F$ can be converted to integers, without scaling.
\begin{lem}\label{lem:conversion}
	Given $(F,H)$,
	there exist
	$F'\in\Z^{\nnn'\times\nnn'}$,
	$\nnn'\le \nnn$,
	$H'\in\R^{\mmm\times\nnn'}$, $\TTT\in\R^{\nnn'\times\nnn}$,
	and $\RRR\in\R^{\nnn'\times\mmm}$
	where $\mathrm{rank}(\TTT)=\nnn'$,
	such that
	$( F' + \RRR H')\TTT = \TTT F$ and $H'\TTT = H$.
\end{lem}

{\it Proof:}
Note that, in \eqref{eq:controller_transformed}, the projection
$z_1=W_1 x$
is defined such that $F_{11}W_1 = W_1 F$ and $H_1 W_1 = H$,
as it guarantees
$ F_{11}(W_1 x)=F_{11} z_1 = W_1 (Fx)$ and $H_1 (W_1 x) = H_1 z_1 = Hx$ for every $x\in\R^\nnn$.
Then,
the existence is guaranteed with
$H'= H_1T_1^{-1}$,
$\TTT = T_1 W_1$, $\RRR= T_1R_1$,
and $ F' = T_1(F_{11}-R_1 H_1)T_1^{-1}$.
It completes the proof.
\hfill $\blacksquare$

\begin{rem}\label{rem:reduced}
Note that, if there is a subset of outputs $u(t)=\{u_i(t)\}_{i=1}^{\mmm}$ from which the system \eqref{eq:controller_transformed_reduced} is observable,
the number of columns for the matrix $\RRR$ in Lemma~\ref{lem:conversion} can be reduced.
For the benefit of such cases, see Remark~\ref{rem:output_decryption}.
\end{rem}

Now,
with the matrices $\TTT$ and $\RRR$ obtained in Lemma~\ref{lem:conversion},
it can be seen how
the controller \eqref{eq:controller_given}
is converted to have the state matrix as integers.
First,
by multiplying both sides of \eqref{eq:controller_given_state} by $\TTT$ from the left,
it yields
\begin{equation}\label{eq:multiplying_T}
\TTT x(t+1) = (F'+\RRR H')\TTT x(t)+\TTT Gy(t) + \TTT P r(t) + \TTT e_x(t).
\end{equation}
Hence,
with respect to a new variable $z:=\TTT x\in\R^{\nnn'}$ for the state,
the converted controller is obtained as the form
\begin{align}\label{eq:controller_converted}
\begin{split}
z(t+1)
&= F' z(t) +(\TTT G -\RRR J)y(t)
\\
&\quad
+(\TTT P -\RRR Q)r(t) +\RRR u(t)+e_z(t)
\\
u(t) &= H'z(t) + Jy(t) + Qr(t) + e_u(t),\\
z(0) &= \TTT x_0 + e_0',
\end{split}
\end{align}
where
it is clear that
the state matrix is converted to integers, as $F'\in\Z^{\nnn'\times \nnn'}$, and
$e_z(t)=\TTT e_x(t) - \RRR e_u(t)$ and $e_0'=\TTT e_0$ denote the perturbations with respect to the state $z(t)$.

For the rest of this subsection,
we consider the stability of closed-loop system,
when the controller \eqref{eq:controller_given} is replaced with \eqref{eq:controller_converted}.
First,
the following proposition states that
the closed-loop is stable with respect to the perturbations $\{e_z(t),e_u(t),e_0'\}$.

\begin{prop}\label{prop:asm1}
	Consider the closed-loop of \eqref{eq:plant} and \eqref{eq:controller_converted}.
Under Assumption~\ref{asm:stability}, there exists $\theta(\epsilon)>0$ such that
if
$\|e_0'\|\le \theta(\epsilon)$,
$\|e_z(t)\|\le \theta(\epsilon)$,
and
$\|e_u(t)\|\le \theta(\epsilon)$
for all $t\in\N_0$,
then
	\begin{align}\label{eq:asm1_converted}
	\|z(t)-\TTT x'(t)\|\le\|\TTT\|\epsilon,~~~\quad\|u(t)-u'(t)\|\le\epsilon,
	\end{align}
and \eqref{eq:asm2.1}
hold,
for all $\ttt\ge 0$ and
$t\in\N_0$,
respectively.
\end{prop}

{\it Proof:}
With  $\TTT'\in\R^{\nnn\times \nnn'}$
such that $\TTT\cdot \TTT'= I_{\nnn'}$, we define
\begin{equation}\label{eq:theta}
\theta(\epsilon):=
\frac{\eta(\epsilon)}{\max(1,\|\TTT'\|(1+\|\RRR\|))}.
\end{equation}
And,
let $\|e_0'\|\le \theta(\epsilon)$, $\|e_z(t)\|\le \theta(\epsilon)$, and $\|e_u(t)\|\le \theta(\epsilon)$, for all $t\in\N_0$.
Consider \eqref{eq:controller_given} as an auxiliary system, with
$e_0=\TTT' e_0'$ and
$e_x(t)=\TTT'(e_z(t)+\RRR e_u(t))$.
As $e_0'=\TTT e_0 $ and $e_z(t) = \TTT e_x(t) - \RRR e_u(t)$,
it follows, from
\eqref{eq:multiplying_T} and \eqref{eq:controller_converted}, that
it satisfies
$z(t)=\TTT x(t)$ for all $t\in\N_0$,
and \eqref{eq:controller_given} and \eqref{eq:controller_converted} have the output $u(t)$ as the same, as well.
As $\theta(\epsilon)$ is defined to guarantee $\|e_0\|\le \eta(\epsilon)$, $\|e_x(t)\|\le \eta(\epsilon)$, and $\|e_u(t)\|\le \eta(\epsilon)$ for all $t$,
Assumption~\ref{asm:stability}
completes the proof,
where the inequality $\|x(t)-x'(t)\|\le\epsilon$ implies that $\|z(t) - \TTT x'(t) \|\le \|\TTT\|\epsilon$.\hfill$\blacksquare$

Finally, the following proposition considers
the effect of perturbations
in \eqref{eq:controller_converted},
when their sizes
may depend on the sizes of $y(t)$,
$r(t)$, $u(t)$, and $z(t)$.
It will be used for
performance analysis of the proposed controllers, in the next subsections.

\begin{prop}\label{prop:eta_linear}
	Consider the closed-loop of \eqref{eq:plant} and \eqref{eq:controller_converted}. Suppose that
	Assumption~\ref{asm:stability} holds,
	$\|e_0'\|\le \theta(\epsilon)$,
	and there exist
	non-negative constants $\theta_z^1$, $\theta_z^2$, $\theta_u^1$, and $\theta_u^2$ such that
\begin{align}\label{eq:prop5}
&\| e_z(t) \|\le \theta_z^1\| [y(t);r(t);u(t);z(t)]\|+\theta_z^2,\notag\\
&\| e_u(t)\|\le \theta_u^1\| [y(t);r(t);z(t)]\|+\theta_u^2,\\
&\theta(\epsilon)\ge \|[\max\{\|\TTT\|,1\}(M+\epsilon)\theta_z^1+\theta_z^2; (M+\epsilon)\theta_u^1+ \theta_u^2] \|.\notag
\end{align}
	Then,
	\eqref{eq:asm2.1} holds for $\ttt\ge 0$,
	and
	\eqref{eq:asm1_converted} holds for
	$t\in\N_0$.
	\end{prop}

{\it Proof:}
At $t=0$,
since $\|e'_0\|\le \theta(\epsilon)$,
Proposition~\ref{prop:asm1} implies
$\|z(0)-\TTT x'(0)\| \le \|\TTT\|\epsilon$ by causality,
and 
$\| z(0)\|\le \|\TTT\|(M+\epsilon)$ by \eqref{eq:bound}.
And,
since
$\|y(0)\|=\|y'(0)\|\le M$ and $\|r(0)\|\le M$,
it follows that
$\|e_u(0)\|\le \theta(\epsilon)$ by \eqref{eq:prop5},
$ \|u(0) - u'(0) \|\le \epsilon $ by Proposition~\ref{prop:asm1} at $t=0$,
and
$\| u(0)\|\le M+\epsilon$ by \eqref{eq:bound}. Hence, $\|e_z(0)\|\le \theta(\epsilon)$ by \eqref{eq:prop5}, sequentially.
Now, suppose
$\|e_z(t)\|\le\theta(\epsilon)$ and $\|e_u(t)\|\le\theta(\epsilon)$
for $t=0,1,...,\tau$.
By Proposition~\ref{prop:asm1} and causality,
\eqref{eq:asm2.1} holds for $0\le \ttt<(\tau+1)\cdot T_s$,
$\|z(t) - \TTT x'(t) \|\le \|\TTT\|\epsilon$ holds for $0\le t \le \tau+1$,
and $\|u(t) - u'(t) \|\le \epsilon$ holds for $0\le t \le \tau$.
Since
$\|z(\tau+1)\|\le \|\TTT\|(M+\epsilon)$,
$\| r(\tau+1)\|\le M$,
and $\|y(\tau+1)\|=\|y_p(\tau\cdot T_s+\tau)\|\le M+\epsilon$ by continuity of \eqref{eq:plant},
it follows that
$\|e_u(\tau+1)\|\le\theta(\epsilon)$ by \eqref{eq:prop5},
and $\|u(\tau+1)\|\le M+\epsilon$ by Proposition~\ref{prop:asm1} and \eqref{eq:bound}. Thus,
$\|e_z(\tau+1)\|\le\theta(\epsilon)$ by \eqref{eq:prop5}.
By induction,
$\|e_z(t)\|\le\theta(\epsilon)$ and $\|e_u(t)\|\le\theta(\epsilon)$ for all $t\in\N_0$.
It concludes that \eqref{eq:asm2.1} holds for all $\ttt\ge 0$, and \eqref{eq:asm1_converted} holds for all $t\in\N_0$.\hfill$\blacksquare$

\subsection{Conversion to System over $\Z_q$}\label{subsec:system}

Since the cryptosystem described in Section~\ref{subsec:homomorphic} considers the set $\Z_q=\{0,1,\cdots,q-1\}$ as the space of plaintexts, and considers modular arithmetic over $\Z_q$,
in this section,
we convert the controller \eqref{eq:controller_converted} to operate over $\Z_q$.
And, we show how to choose the modulus $q\in\N$,
in a non-conservative way.

First,
we consider implementation of \eqref{eq:controller_converted} over quantized integers, as well as its performance.
Since the matrices $\TTT$, $\RRR$, and $H'$ may consist of irrational numbers, we let the irrational matrices in \eqref{eq:controller_converted} be scaled and truncated as
$$
\ol G':= \left\lceil{\frac{\TTT G - \RRR J}{\sss_1}}\right\rfloor,~~
\ol P':= \left\lceil\frac{\TTT P - \RRR Q}{\sss_1}\right\rfloor,~~
\ol \RRR:= \left\lceil\frac{\RRR}{\sss_1}\right\rfloor,
$$
and $ \ol H':= \lceil H' / \sss_2\rfloor$,
so that they are stored as integers.
Then, the controller \eqref{eq:controller_converted} can be implemented as the same as \eqref{eq:quantized_rev},
as
\begin{align}\label{eq:controller_converted_integer}
\ol z(t+1) &=  F'\ol z(t) + \ol G' \ol y(t) + \ol P' \ol r(t) +\ol \RRR {\ol u}' (t), \\
{\ol u}(t)&= \ol H' \ol z(t) + \ol J \ol y(t) + \ol Q \ol r(t),~~\ol u'(t):= \lceil\sss_1\sss_2\cdot {\ol u} (t)\rfloor,\notag
\end{align}
where $\ol z(t)\in\Z^{\nnn'}$ is the state, with $\ol z(0) = \lceil(\TTT x_0)/(\rrr_1\sss_1) \rfloor$, and $\ol u(t)\in\Z^\mmm$ is the output.

Note that, as the same as in \eqref{eq:quantized_rev_controller} with \eqref{eq:input_quantization},
the state $\ol z(t)$ should be of scale $1/(\rrr_1\sss_1)$,
and the output $\ol u(t)$ should be of scale $1/(\rrr_1\sss_1\sss_2)$,
and the inputs $\ol y(t)$, $\ol r(t)$, and $\ol u'(t)$ of \eqref{eq:controller_converted_integer} should be of scale $1/\rrr_1$,
so that
the computation for $\ol u'(t)$
divides $\ol u(t)$ by the factor $1/(\sss_1\sss_2)$.
And, note that the converted state matrix $F'$ consists of integers, thanks to Lemma~\ref{lem:conversion}, so that it does not need to be scaled.
The recovery of the plant input $u(t)$ from the output $\ol u(t)$ is defined
as the same as in \eqref{eq:output_quantization}.

As a result, the following proposition shows that the performance of the converted controller \eqref{eq:controller_converted_integer} over integers is equivalent to
that of
the given controller \eqref{eq:controller_given},
when the parameters $\{1/\rrr_1,1/\rrr_2,1/\sss_1,1/\sss_2\}$ are chosen sufficiently large.

\begin{prop}\label{prop:quantized}
	Under Assumption~\ref{asm:stability},
	there exists
	a continuous function $\beta(\rrr_1,\rrr_2,\sss_1,\sss_2)\in\R$
	vanishing at the origin such that
	if
	$\beta(\rrr_1,\rrr_2,\sss_1,\sss_2)\le \theta(\epsilon)$,
	then
	the controller
\eqref{eq:controller_converted_integer} with \eqref{eq:output_quantization}
	ensures that $\|u(t)-u'(t)\|\le \epsilon$ holds,
for all	$t\in\N_0$.
\end{prop}

{\it Proof:}
We identify the system \eqref{eq:controller_converted_integer} with the system \eqref{eq:controller_converted},
by $z(t) = \rrr_1\sss_1\ol z(t)$ and $u(t) = \QQQ^u(\ol u(t))$,
where the perturbations $\{e_z(t),e_u(t),e_0'\}$ in \eqref{eq:controller_converted} are determined as

\vspace{-4mm}
{\small
\begin{align*}
e_z(t) &= \rrr_1\sss_1\left\lceil \frac{\TTT G - \RRR J}{\sss_1}\right\rfloor \left\lceil \frac{y(t)}{\rrr_1}\right\rfloor - (\TTT G - \RRR J) y(t)\\
&\quad + \rrr_1\sss_1\left\lceil \frac{\TTT P - \RRR Q}{\sss_1}\right\rfloor \left\lceil \frac{r(t)}{\rrr_1}\right\rfloor - (\TTT P - \RRR Q) r(t)\\
&\quad + \rrr_1\sss_1\left\lceil \frac{\RRR}{\sss_1}\right\rfloor \lceil \sss_1\sss_2\ol u(t)\rfloor - \RRR \cdot \rrr_1\sss_1\sss_2 \ol u(t),\\
e_u(t) &= \QQQ^u(\ol u(t)) - \rrr_1\sss_1\sss_2\ol u(t) + \sss_2\left\lceil \frac{H'}{\sss_2}\right\rfloor z(t) - H'z(t)\\
&\quad+\rrr_1 J \left\lceil \frac{y(t)}{\rrr_1}\right\rfloor - Jy(t)+\rrr_1 Q \left\lceil \frac{r(t)}{\rrr_1}\right\rfloor - Qr(t),
\end{align*}}
and $e_0'=\rrr_1\sss_1\lceil(\TTT x_0)/(\rrr_1\sss_1)\rfloor - \TTT x_0 $.
From \eqref{eq:output_quantization},
note that $\|u(t) - \rrr_1\sss_1\sss_2\ol u(t) \|\le \rrr_2/2$.
Then, from the fact that
$$ \left\|\rrr_1\sss_1 \left\lceil \frac{S}{\sss_1} \right\rfloor \left\lceil  \frac{w}{\rrr_1}  \right\rfloor - Sw\right\|
\le \frac{l_2}{2}\|w\|\sss_1 + \frac{\|S\|}{2}\rrr_1 + \frac{l_2}{4}\rrr_1\sss_1 $$
holds for $S\in\R^{l_1\times l_2}$ and $w\in\R^{l_2}$,
it follows that

\vspace{-3mm}
{\small
\begin{align}
\|e_z(t)\|&\le \frac{l_2}{2}\|w(t)\|\sss_1 + \frac{\|S\|}{2}\rrr_1+\frac{l_2}{4}\rrr_1\sss_1\notag
\\
&\le \frac{l_2}{2}\|[y(t);r(t);u(t)]\|\sss_1+ \frac{\|S\|}{2}\rrr_1\!+\!
\frac{l_2}{4}(\rrr_1\!+\!\rrr_2)\sss_1\notag
\\
&=: \beta_z(\| [y(t);r(t);u(t)]\|,\rrr_1,\rrr_2,\sss_1,\sss_2)\label{eq:e_z}
\end{align}}
where $S= [\TTT G - \RRR J , \TTT P - \RRR Q, \RRR]$,
$l_2 = \ppp + \mmm + \qqq$, and $w(t) = [y(t);r(t);\rrr_1\sss_1\sss_2 \ol u(t)]$.
And, we further have
\begin{align}
\begin{split}\label{eq:e_u}
\|e_u(t)\| &\le \frac{1}{2}\rrr_2+ \frac{\nnn'}{2}\|z(t)\|\sss_2 + \frac{\|[J,Q]\|}{2}\rrr_1\\
&=:\beta_u(\|z(t)\|,\rrr_1,\rrr_2,\sss_2),
\end{split}
\end{align}
and $\|e_0'\|\le (\rrr_1\sss_1)/2$.
With these terms, we define
\begin{equation}\label{eq:beta}
\beta(\rrr_1,\rrr_2,\sss_1,\sss_2)
:={\small\left\|\begin{bmatrix}
\beta_z(M+\epsilon,\rrr_1,\rrr_2,\sss_1,\sss_2) \\
\beta_u(\|\TTT\|(M+\epsilon),\rrr_1,\rrr_2,\sss_2)\\
\frac{1}{2}\rrr_1\sss_1
\end{bmatrix}\right\|},
\end{equation}
which is clearly continuous and vanishes at the origin.
Choose $(\rrr_1,\rrr_2,\sss_1,\sss_2)$ such that
$\beta(\rrr_1,\rrr_2,\sss_1,\sss_2)\le \theta(\epsilon)$.
Then,
with \eqref{eq:e_z} and \eqref{eq:e_u},
Proposition~\ref{prop:eta_linear} completes the proof.\hfill $\blacksquare$

Now, for encryption, we implement \eqref{eq:controller_converted_integer} over the space $\Z_q$, with modular arithmetic.
Following the consideration in Remark~\ref{rem:injected_errors},
which is to deal with the injected errors during encryption,
we propose that all the quantized signals to be encrypted should be first scaled by an additional parameter\footnote{For the case of utilizing cryptosystems without error injection, i.e., cryptosystems satisfying H1 with $\Delta_\Enc=0$, H2, and H3 or H4, there is no need to introduce
	the parameter $1/\LLL$, i.e., it can be assumed that $1/\LLL=1$.
} $1/\LLL\in\N$.
For this,
let the signals of \eqref{eq:controller_converted_integer} be scaled by $1/\LLL$,
as
\begin{align}\label{eq:controller_converted_scaled_integer}
\begin{split}
\tilde z(t+1) &=  F'\tilde z(t) + \ol G' \!\cdot\!\frac{\ol y(t)}{\LLL} + \ol P'
\!\cdot\! \frac{\ol r(t)}{\LLL} +\ol \RRR \!\cdot\!\frac{{\ol u}' (t)}{\LLL}, \\
{\tilde u}(t)&= \ol H' \tilde z(t) + \ol J \!\cdot\!\frac{\ol y(t)}{\LLL} + \ol Q \!\cdot\!
\frac{\ol r(t)}{\LLL},~\tilde z(0) =\frac{\ol z(0)}{\LLL},
\end{split}
\end{align}
where the factor $1/\LLL$ does not affect the performance of \eqref{eq:controller_converted_integer}, since $\ol z(t)= \LLL\cdot  \tilde z(t)$ and $\ol u(t) = \LLL\cdot \tilde u(t)$, $\forall t\in\N_0$.

Then, we convert \eqref{eq:controller_converted_scaled_integer} to operate over $\Z_q$.
By taking the modulo operation,
let the system \eqref{eq:controller_converted_scaled_integer} over $\Z$
be projected as
\begin{align}\label{eq:controller_converted_mod}
\tilde \zzz(t+1) &=  F'\tilde \zzz(t) + \ol G' \!\cdot\!\frac{\ol y(t)}{\LLL} + \ol P'
\!\cdot\! \frac{\ol r(t)}{\LLL} +\ol \RRR \!\cdot\!\frac{{\ol u}' (t)}{\LLL} \!\mod q,\notag  \\
{\tilde \uuu}(t)&= \ol H' \tilde \zzz(t) + \ol J \!\cdot\!\frac{\ol y(t)}{\LLL} + \ol Q \!\cdot\!
\frac{\ol r(t)}{\LLL} \mod q,
\end{align}
in which $\tilde \zzz(0) = \ol z(0)/\LLL \mod q$,
so that it operates with modular addition and multiplication over the space $\Z_q$.
Note that the matrices and the scaled inputs in
\eqref{eq:controller_converted_mod} can be regarded as elements of $\Z_q$, by taking the modulo operation, as well.

In the remaining,
we find a lower bound for the modulus $q\in\N$ such that the performance of \eqref{eq:controller_converted_mod} can be equivalent to that of \eqref{eq:controller_converted_scaled_integer},
with which we state the theorem of this subsection.
We propose that it is enough to have the modulus $q$ cover the range set of the controller output $\tilde u(t)$ only,
as it is the only signal to be fed back to the plant.
To this end,
let us suppose that
the output $\tilde u(t)=\{\tilde u_i(t)\}_{i=1}^{\mmm}$ of \eqref{eq:controller_converted_scaled_integer} be bounded as
\begin{equation}\label{eq:tilde_bound}
\tilde u_i^{\min} \le \tilde  u_i(t) \le \tilde u_i^{\max},\quad\forall t\in\N_0,
\quad
\forall
i=1,\cdots,\mmm,
\end{equation}
with some integers\footnote{We defer the definition
of $\{\tilde u_i^{\min},\tilde u_i^{\max}\}_{i=1}^{\mmm}$, which will be defined in \eqref{eq:input_integer_range}, with the knowledge of the bound for the output of the given model \eqref{eq:controller_given}.
}
$\{\tilde u_i^{\min},\tilde u_i^{\max}\}_{i=1}^{\mmm}$.
And, let us define
\begin{equation*}
U:=\{v\in\Z^\mmm: v= w+\col\{\tilde u_i^{\min}\}_{i=1}^{\mmm} , w\in\Z_q^\mmm \},
\end{equation*}
which is a parallel transport of the set $\Z_q^\mmm$ along the vector $\col\{\tilde u_i^{\min}\}_{i=1}^{\mmm}\in\Z^\mmm$,
and choose the modulus $q$
to satisfy
\begin{equation}\label{eq:N}
q \ge \max_{i=1,...,\mmm}\{\tilde u_{i}^{\max}-\tilde u_{i}^{\min}+1\}, 
\end{equation}
so that
the set $U$ covers the range \eqref{eq:tilde_bound}.
Then,
by taking a ``biased'' modulo operation for the output of \eqref{eq:controller_converted_mod}, defined as\footnote{Note that the operation 
$\mathrm{mod}~U$
projects a vector in $\Z^\mmm$ into the set $U$.}
\begin{equation}\label{eq:biased}
\tilde \uuu(t)\mod U := \tilde \uuu(t) - \left\lfloor{\small \frac{\tilde \uuu(t)- 
	\col\{\tilde u_i^{\min}\}_{i=1}^{\mmm}
 }{q}}\right\rfloor q,
\end{equation}
the following lemma shows that the system \eqref{eq:controller_converted_mod} over $\Z_q$ can yield the same output $\tilde u(t)$ of the system \eqref{eq:controller_converted_scaled_integer}, as long as the output $\tilde u(t)$ does not exceed the bounds $\{\tilde u_i^{\min},\tilde u_i^{\max}\}_{i=1}^{\mmm}$.

\begin{lem}\label{lem:mod}
	Suppose that two systems \eqref{eq:controller_converted_scaled_integer} and \eqref{eq:controller_converted_mod} share the same signals $\ol y(t)$, $\ol r(t)$, and $\ol u'(t)$ as external inputs.
	If \eqref{eq:tilde_bound} and \eqref{eq:N} hold, then $\tilde u(t) = \tilde \uuu(t)\mod U$, $\forall t\in\N_0$.
\end{lem}

{\it Proof:}
By construction, it is obvious that
$\tilde \zzz(t) = \tilde z(t)\mod q$, and $\tilde \uuu(t)=\tilde u(t)\mod q$, for all $t\in\N_0$,
since \eqref{eq:controller_converted_scaled_integer} and \eqref{eq:controller_converted_mod} share the same input.
Since
$\tilde \uuu(t) = \tilde u(t) + k\cdot q$ with some $k\in\Z^\mmm$,
and \eqref{eq:tilde_bound} and \eqref{eq:N} imply $0\le \tilde u_i(t) - \tilde u_i^{\min }<q$
for each $ i=1,\cdots,\mmm$,
it follows that
$
\tilde \uuu(t) \mod U
= \tilde u(t),
$
from \eqref{eq:biased}.
It completes the proof.\hfill$\blacksquare$

And, the idea is extended to consider the performance of the controller \eqref{eq:controller_converted_mod}, in the closed-loop system.
Let
the input $u(t)$ of the plant \eqref{eq:plant}
be recovered 
from the output $\tilde \uuu(t)$ of \eqref{eq:controller_converted_mod},
by
\begin{equation}\label{eq:output_recovery}
u(t) = \QQQ^u(\LLL\cdot (\tilde \uuu(t)\mod U)),
\end{equation}
where
$\QQQ^u(\cdot)$
is the same quantization function defined in \eqref{eq:output_quantization}.
Then, the following lemma ensures that
the performance of \eqref{eq:controller_converted_mod} over $\Z_q$ is identically the same with that of \eqref{eq:controller_converted_scaled_integer} over $\Z$,
in the closed-loop,
as long as the output $\tilde u(t)$ of \eqref{eq:controller_converted_scaled_integer} does not exceed its bound,
and
the modulus $q$ covers its range.
\begin{lem}\label{lem:mod_closed}
	Consider the closed-loop of \eqref{eq:plant} and \eqref{eq:controller_converted_scaled_integer} with 
	$u(t)=\QQQ^u(\LLL\cdot \tilde u(t))$,
	and that of \eqref{eq:plant} and \eqref{eq:controller_converted_mod} with \eqref{eq:output_recovery}.
	Assuming that \eqref{eq:tilde_bound} and \eqref{eq:N} hold,
	they satisfy $\tilde u(t) = \tilde \uuu(t)\mod U$, for all $t\in\N_0$.
\end{lem}

{\it Proof:}
For the closed-loop of \eqref{eq:plant} and \eqref{eq:controller_converted_scaled_integer},
let
$u^a(t)$,  $x_p^a(\ttt)$, and $\ol y^a(t)$
denote
the signals $u(t)$, $x_p(\ttt)$, and $\ol y(t)$, respectively.
First,
it is obvious that
$\tilde \zzz(0)= \tilde z(0)
~\mathrm{mod}\,q
$
and
$x_p(0) = x_p^a(0)$.
Now, suppose that $\tilde \zzz(t) = \tilde z(t)\mod q$ for $t = 0,\cdots,\tau$,
and
$x_p(\ttt)=x_p^a(\ttt)$ for $0\le \ttt \le\tau T_s$.
It implies $\ol y(\tau)=\ol y^a(\tau)$.
Since the two closed-loops have the reference $\ol r(t)$ as the same,
and \eqref{eq:tilde_bound} holds,
it follows that $\tilde u(\tau) = \tilde \uuu(\tau) \mod U$,
by Lemma~\ref{lem:mod} with causality.
It follows that $u(\tau)=u^a(\tau)$,
and
$x_p(\ttt)= x_p^a(\ttt)$ for $\tau T_s\le \ttt < (\tau+1) T_s$,
and
$x_p((\tau+1) T_s)= x_p^a((\tau+1) T_s)$,
by continuity.
Since $u(\tau)=u^a(\tau)$,
the two systems have the input $\ol u'(t)$ as the same at $t=\tau$, so it follows that
$\tilde \zzz(t) = \tilde z(t) \mod q$.
Then, by induction,
it satisfies
$x_p(\ttt)=x_p^a(\ttt)$, $\forall \ttt\ge 0$,
and
$\tilde \zzz(t) = \tilde z(t) \mod q$, $\forall t\in\N_0$,
so that $\tilde u(t) = \tilde \uuu(t)\mod U$, $\forall t\in\N_0$.
It completes the proof.\hfill$\blacksquare$

Now,
we
choose the parameters $\{\tilde u_i^{\min},\tilde u_i^{\max}\}_{i=1}^{\mmm}$ for the controller \eqref{eq:controller_converted_mod},
so that
the output $\tilde u(t)$ of \eqref{eq:controller_converted_scaled_integer} satisfies
\eqref{eq:tilde_bound}.
From the boundedness of the output $u'(t)=\col\{u'_i(t)\}_{i=1}^{\mmm}\in\R^\mmm$ of \eqref{eq:controller_given}, as the perturbation free case,
let $u_i^{\min}\in\R$ and $u_i^{\max}\in\R$, be constants
such that
the signal
$u_i'(t)$
is bounded as $u_i^{\min} \le u'_i(t) \le u_i^{\max}$,
for all $t\in\N_0$.
With this, we define
\begin{align}\label{eq:input_integer_range}
\tilde u_{i}^{\min}:= 
{\small\left\lfloor \frac{u_i^{\min}-\epsilon-\frac{\rrr_2}{2}}{\LLL\rrr_1\sss_1\sss_2}\right\rfloor},\,
\tilde u_{i}^{\max}:={\small\left\lceil \frac{u_i^{\max} + \epsilon+\frac{\rrr_2}{2}}{\LLL\rrr_1\sss_1\sss_2}\right\rceil},
\end{align}
for each $i=1,\cdots,\mmm$,
where the factor $1/(\LLL\rrr_1\sss_2\sss_2)$ considers that the output $\tilde u(t)$ has the value of $u(t)/(\LLL\rrr_1\sss_2\sss_2)$,
and the term $\epsilon+\rrr_2/2$ is for the margin of error due to quantization.

Finally, the following theorem states that
the performance of the converted controller \eqref{eq:controller_converted_mod} over $\Z_q$
is equivalent to the given controller \eqref{eq:controller_given}, when the parameters $\{\rrr_1,\rrr_2,\sss_1,\sss_2\}$ are chosen sufficiently large, as the same as in Proposition~\ref{prop:quantized}.

\begin{thm}\label{thm:mod}
Suppose that
Assumption~\ref{asm:stability} holds, and $\beta(\rrr_1,\rrr_2,\sss_1,\sss_2)\le \theta(\epsilon)$.
	Then,
	the controller
	\eqref{eq:controller_converted_mod} over $\Z_q$,
	with \eqref{eq:output_recovery},
	guarantees that
	$\|
	u(t)
	-u'(t)\|\le \epsilon$,
	$
	\forall
	t\in\N_0$.
\end{thm}

{\it Proof:}
As an auxiliary system,
consider a closed-loop of a copy of \eqref{eq:plant} and \eqref{eq:controller_converted_scaled_integer} with $u(t) = \QQQ^u(\LLL\tilde u(t))$.
Then,
by Proposition~\ref{prop:quantized},
$\|\QQQ^u(\LLL\tilde u(t))-u'(t)\| \le \epsilon$, $\forall t\in\N_0$.
Since
\eqref{eq:output_quantization} implies $\|\QQQ^u(\LLL \tilde u(t)) - \LLL\rrr_1\sss_2\sss_2 \tilde u(t)  \|\le \rrr_2/2$,
it implies $\|\LLL\rrr_1\sss_2\sss_2 \tilde u(t) - u'(t) \|\le \epsilon + \rrr_2/2$, and hence the design \eqref{eq:input_integer_range} guarantees \eqref{eq:tilde_bound}.
Then, Lemma~\ref{lem:mod_closed} completes the proof.\hfill$\blacksquare$

\begin{rem}\label{rem:cut}
	In terms of the state of the designed controller \eqref{eq:controller_converted_mod},
	Lemma~\ref{lem:mod} guarantees no more than the property $\tilde \zzz(t)=\tilde z(t)\mod q$.
	Indeed,
	if we write $\tilde z(t) = \tilde z_1(t)\cdot q + \tilde z_2(t)$,
	with some
	$\tilde z_1(t)\in\Z^{\nnn'}$ and
	$\tilde z_2(t)\in\Z_q^{\nnn'}$,
	the portion $\tilde z_1(t)$ (the higher bits of the state $\tilde z(t)$ of \eqref{eq:controller_converted_scaled_integer}) is cut off in \eqref{eq:controller_converted_mod}.
	This is because
	the modulus $q$ is chosen to satisfy \eqref{eq:N} only, i.e., to cover the range of output $\tilde u(t)$ only,
	to reduce the conservatism.
	Nonetheless,
	as long as Lemma~\ref{lem:mod_closed} shows that the performances of
	\eqref{eq:controller_converted_scaled_integer} and \eqref{eq:controller_converted_mod} are identically the same unless there is an overflow for the outputs,
	Theorem~\ref{thm:mod}
	guarantees that
	the performance of 
	\eqref{eq:controller_converted_mod}
	is equivalent to  the given controller \eqref{eq:controller_given},
	despite cutting off the higher bits of the state $\tilde z(t)$.
\end{rem}

\begin{rem}
	In the previous results on encrypted control, as in
	\cite{Farokhi17CEP}, \cite{Darup19CSL}, or \cite{Cheon18},
	the modulus $q\in\N$ has been commonly chosen conservatively large,
	to cover all the ranges of the input, state, and output of the controller.
	Indeed,
	by replacing the modulo operations in \eqref{eq:controller_converted_mod}
	and
	\eqref{eq:biased}
	with the (component-wise) projection
	onto the space $\bar \Z_q:=\{i\in\Z: -\frac{q}{2}\le i < \frac{q}{2}\}$,
	and by increasing $q$ for enlarging the space $\Z_q$,
	it is trivial that the operation of \eqref{eq:controller_converted_mod} can be identical with that of \eqref{eq:controller_converted_scaled_integer},
	in which the projection operation does nothing about the values.
	In contrast,
	the proposed criterion
	based on the idea of Lemma~\ref{lem:mod},
	which chooses the modulus $q$ to cover the range of the output only,
	can be seen as a less conservative way.	
	The method of \eqref{eq:controller_converted_mod}
	also suggests that
	there is no need of
	defining the space of quantized numbers as the space $\bar\Z_q$,
	for dealing with both positive integers and negative integers in
	\eqref{eq:controller_converted_scaled_integer}.		
\end{rem}

Now, it is straightforward to implement \eqref{eq:controller_converted_mod}
over encrypted data,
by exploiting the properties H1--H4 of the cryptosystem,
since the operation of \eqref{eq:controller_converted_mod}
comprises only modular addition and multiplication over $\Z_q$.
In the next section,
the main result of encrypted dynamic controller that operates for an infinite time horizon is presented,
where the effect of injected errors during encryption is analyzed.

\subsection{Controller over Encrypted Data}

The conversion of the given controller \eqref{eq:controller_given} to the system \eqref{eq:controller_converted_mod} over $\Z_q$ directly allows the operation over encrypted data.
Let the plant output $\ol y(t)\in\Z^\mmm$ and the reference $\ol r(t)\in\Z^\qqq$, as quantized signals defined in \eqref{eq:input_quantization},
be encrypted as
\begin{align*}
\yy(t) &:= \Enc\left(\frac{\ol y(t)}{\LLL}~\mathrm{mod}\,q\right),\quad
\!\rr(t) := \Enc\left(\frac{\ol r(t)}{\LLL}~\mathrm{mod}\,q\right),
\end{align*}
with the scale factor $1/\LLL\in\N$.
And, to conceal
the matrices
in
\eqref{eq:controller_converted_mod} or \eqref{eq:controller_converted}
as well, they are encrypted as ``multipliers,''
a priori,
with the algorithm $\Enc'$ introduced in Section~\ref{subsec:homomorphic},
as
\begin{align*}
\FF'&:= \Enc'(F' \mod q),\quad 
\GG':= \Enc'(\ol G' \mod q),\notag\\
\HH'&:= \Enc'(\ol H' \mod q),\quad~\!~
\JJ:= \Enc'(\ol J \mod q),\notag\\
\PP'&:= \Enc'(\ol P' \mod q),\quad~
\QQ:= \Enc'(\ol Q \mod q),\notag\\
\RR &:= \Enc'(\ol \RRR \mod q).
\end{align*}

With these encrypted signals and matrices,
the proposed encrypted dynamic controller over $\CC$ is constructed, as\footnote{Note that
	the operations $(+,\,\cdot\,)$ in \eqref{eq:controller_encrypted}, over $\CC$, are defined in Section~\ref{subsec:homomorphic},
	with the algorithms $\Add$ and $\Mult$ of H2 and H4, respectively.
}
\begin{align}\label{eq:controller_encrypted}
\zz(t+1) &= \FF'\cdot \zz(t) + \GG'\cdot \yy(t)+ \PP'\cdot \rr(t)+ \RR\cdot  \uu'(t)\notag\\
\uu(t) &= \HH'\cdot\zz(t)+\JJ\cdot\yy(t)+\QQ\cdot\rr(t),
\end{align}
with $\zz(0) = \Enc(\ol z(0)/\LLL\mod q)$,
where
$\zz(t)\in\CC^{\nnn'}$ is the state,  $\uu(t)\in\CC^\mmm$ is the output,
and where we let the output $\uu(t)$ be decrypted at the plant, as
$$ \tilde u_e(t):=\Dec(\uu(t))\mod U,$$
so that the plant input $u(t)\in\R^\mmm$, and the signal $\uu'(t)\in\CC^\mmm$ as an external input of \eqref{eq:controller_encrypted},
are obtained as\footnote{In practice, in \eqref{eq:controller_encrypted_u},
	the encryption rule for $\uu'(t)$ can be replaced with
	$\uu'(t) = \Enc(\lceil \sss_1\sss_2\cdot\tilde u_e(t)\rfloor\mod q)$,
	to preserve more lower bits of $\tilde u_e(t)$.}
\begin{equation}\label{eq:controller_encrypted_u}
u(t) = \QQQ^u (\LLL \cdot \tilde u_e(t)),~ \uu'(t)=
\Enc\left(\frac{\lceil \LLL\sss_1\sss_2\cdot \tilde u_e(t)\rfloor}{\LLL}
~\mathrm{mod}\,q
\right),
\end{equation}
respectively,
which are analogous to \eqref{eq:output_recovery} and \eqref{eq:controller_converted_integer}.

Finally,
the following main theorem states that given any controller \eqref{eq:controller_given}, it can be encrypted
as \eqref{eq:controller_encrypted} with \eqref{eq:controller_encrypted_u},
to operate
for an infinite time horizon,
where the performance error
can be made arbitrarily small
with the choice of parameters.

\begin{thm}\label{thm:main}
Under Assumption~\ref{asm:stability},
there exists a continuous function $\gamma(\LLL,\rrr_1,\rrr_2,\sss_1,\sss_2)\in\R$ vanishing at the origin such that if $\gamma(\LLL,\rrr_1,\rrr_2,\sss_1,\sss_2) \le \theta(\epsilon)$, then
the encrypted controller \eqref{eq:controller_encrypted}
with \eqref{eq:controller_encrypted_u} 
guarantees that $\|u(t) - u'(t)\|\le \epsilon$, $\forall t\in\N_0$.
\end{thm}

{\it Proof:}
We define $\tilde \zzz_e(t):=\Dec(\zz(t))\in\Z_q^{\nnn'}$
and
$\tilde \uuu_e(t):=\Dec(\uu(t))\in\Z_q^\mmm$.
From
\eqref{eq:MatMult} and H1,
it follows that
\begin{align}\label{eq:controller_encryption_decrypted}
\tilde \zzz_e(t+1)&= F' \tilde \zzz_e(t)+ \ol G' \frac{\ol y(t)}{\LLL} + \ol P' \frac{\ol r(t)}{\LLL}+\ol \RRR \frac{\lceil \LLL\sss_1\sss_2\cdot \tilde u_e(t)\rfloor}{\LLL}\notag\\
&\quad
+\Delta_{1}(t)\mod q\notag\\
\tilde \uuu_e(t)
&= \ol H' \tilde \zzz_e(t)+\ol J \frac{\ol y(t)}{\LLL} +  \ol Q \frac{\ol r(t)}{\LLL} + \Delta_{2}(t) \mod q\notag\\
\tilde \zzz_e(0)&= \frac{1}{\LLL}\left\lceil\frac{\TTT x_0}{\rrr_1\sss_1} \right\rfloor+\Delta_0\mod q
\end{align}
with some $\Delta_1(t)\in\Z^{\nnn'}$, $\Delta_2(t)\in\Z^\mmm$, and $\Delta_0\in\Z^{\nnn'}$ such that
\begin{align}\label{eq:Delta}
\|\Delta_1(t)\|
&\le \left(\frac{\|[\TTT G \!-\! \RRR J , \TTT P \!-\! \RRR Q, \RRR] \|}{\sss_1} + \frac{\ppp+\qqq+\mmm}{2}\right)\Delta_\Enc 
\notag
\\
&\quad
+ (\nnn'+\ppp+\qqq+\mmm)\Delta_\Mult=:\gamma_1(\sss_1),\\
\|\Delta_2(t)\| &\le\frac{\|[J,Q]\|}{\sss_1\sss_2}\Delta_\Enc
 + (\nnn' + \ppp + \qqq)\Delta_\Mult=: \gamma_2(\sss_1,\sss_2),\notag
\end{align}
and
$\|\Delta_0\|\le \Delta_\Enc$.
Now, as an auxiliary system,
we consider closed-loop of a copy of \eqref{eq:plant},
and the system 
\eqref{eq:controller_converted_scaled_integer} perturbed as
\begin{align}\label{eq:controller_encryption_equivalent}
\tilde z^a(t+1)&= F' \tilde z^a(t)+ \ol G' \frac{\ol y(t)}{\LLL} + \ol P' \frac{\ol r(t)}{\LLL} + \ol \RRR \frac{\lceil \LLL\sss_1\sss_2\cdot \tilde u^a(t)\rfloor}{\LLL}\notag \\&\quad+ \Delta_1(t)\notag\\
\tilde u^a(t)
&= \ol H' \tilde z^a(t)+\ol J\frac{\ol y(t)}{\LLL} + \ol Q \frac{\ol r(t)}{\LLL} + \Delta_2(t)\\
\tilde z^a(0)&=\frac{1}{\LLL}\left\lceil\frac{\TTT x_0}{\rrr_1\sss_1} \right\rfloor+\Delta_0,
\notag
\end{align}
with $u(t) = \QQQ^u(\LLL\tilde u^a(t))$.
Then, with the function $\beta$ given from Proposition~\ref{prop:quantized}, we define
\begin{equation}\label{eq:thm1_beta}
\gamma(\LLL,\rrr_1,\rrr_2,\sss_1,\sss_2):= \beta(\rrr_1,\rrr_2,\sss_1,\sss_2)+{\small\left\|\begin{bmatrix}
\LLL\rrr_1\sss_1\cdot \gamma_1(\sss_1)\\
\LLL\rrr_1\sss_1\sss_2\cdot \gamma_2(\sss_1,\sss_2)\\
\LLL\rrr_1\sss_1\Delta_\Enc
\end{bmatrix}\right\|},
\end{equation}
which is continuous and vanishes at the origin.
Since the system \eqref{eq:controller_encryption_equivalent} is equivalent to \eqref{eq:controller_converted_integer},
with the state $\ol z(t)$,
the output $\ol u(t)$,
and the initial state $\ol z(0)$ perturbed by
$\LLL\Delta_1(t)$, $\LLL\Delta_2(t)$, and $\LLL\Delta_0$, respectively,
it can be easily verified that
$\gamma(\LLL,\rrr_1,\rrr_2,\sss_1,\sss_2)\le \theta(\epsilon)$ implies
$\|\QQQ^u(\LLL\tilde u^a(t))-u'(t)\|\le \epsilon$ for all $t\in\N_0$,
as analogous to the proof of
Proposition~\ref{prop:quantized}.
Then,
as the same as in the proof of Theorem~\ref{thm:mod},
it follows that
for each $i=1,\cdots,\mmm$,
the output $\tilde u^a(t)=\{\tilde u_i^a(t)\}_{i=1}^{\mmm}$ is bounded as $\tilde u_i^{\min}\le \tilde u_i^a(t)\le \tilde u_i^{\max}$, for all $t\in\N_0$.
By applying Lemma~\ref{lem:mod_closed} to the controllers \eqref{eq:controller_encryption_decrypted} and \eqref{eq:controller_encryption_equivalent} with their respective plant,
it follows that $\|\QQQ^u(\LLL
\tilde u_e(t)
) -u'(t)\|=\|\QQQ^u(\LLL \tilde u^a(t)) -u'(t)\|\le \epsilon$,  for all $ t\in\N_0$. It completes the proof.
\hfill$\blacksquare$

\begin{rem}\label{rem:L}
	From  \eqref{eq:thm1_beta},
	which shows that
	the performance error of the encrypted controller is due to quantization and the effect of injected errors during encryption,
	we note that $\gamma(0,\rrr_1,\rrr_2,\sss_1,\sss_2)=\beta(\rrr_1,\rrr_2,\sss_1,\sss_2)$.
	It means that
	the latter term of \eqref{eq:thm1_beta}, which is for the effect of injected errors, can be made arbitrarily small, by increase\footnote{
		In practice,
		the size of the constant
		$\Delta_\Mult$ in the function $\gamma$,
		the error growth by multiplication,
		may increase, when the parameter $1/\LLL$ increases.
		Nonetheless,
		the claim of Remark~\ref{rem:L} is still true for such cases. For more details, refer to a following up result \cite{Kim20CDC} of this paper.} of $1/\LLL$.
	In this sense, the performance of the encrypted controller \eqref{eq:controller_encrypted} can be seen as equivalent to that of \eqref{eq:controller_converted_integer}, with sufficiently large $1/\LLL$.
	\end{rem}

\begin{rem}\label{rem:output_decryption}
The output $\uu(t)\in\CC^\mmm$ is required to be re-encrypted and used as the term $\RR\uu'(t)$ in \eqref{eq:controller_encrypted}.
	Compared with the model re-encrypting the state $\xx(t)\in\CC^\nnn$,
	the number of encrypted messages required for re-encryption is changed from $\nnn$ to $\mmm$.
	Thanks to Remark~\ref{rem:reduced},
	it can be further reduced to the minimal number of outputs from which the controller is observable, which is obviously less than or equal to $\nnn$.
\end{rem}

\begin{rem1}
Extending the discussion of Remark~\ref{rem:output_decryption},
it is notable that systems having the state matrix as integers can be encrypted to operate for infinite time horizon, without re-encryption of output.
Indeed,
given the controller \eqref{eq:controller_given} with $F\in\Z^{\nnn\times\nnn}$,
the method of Section~\ref{subsec:system} can be directly applied with $\TTT=I_{\nnn\times\nnn}$ and $\RRR=0_{\nnn\times\mmm}$,
so that the term $\RR\cdot\uu'(t)$ in \eqref{eq:controller_encrypted} is dispensable.
Such cases of controllers which have the state matrix as integers are found,
as follows.

	\begin{itemize}
		\item Finite-Impulse-Response (FIR) controller:
		With the transfer function $C(z) = \sum_{i=0}^n b_{n-i}z^{-i}$, it is realized as
		\begin{align*}
		x(t+1) &= {\small\begin{bmatrix}
		0 & \cdots & 0 & 0\\
		1 & \cdots & 0 & 0\\
		\vdots & \ddots & \vdots & \vdots\\
		0 & \cdots & 1 & 0
		\end{bmatrix}}x(t) + {\small\begin{bmatrix}
		1\\0\\\vdots \\ 0
		\end{bmatrix}} y(t)\\
		u(t) &= \begin{bmatrix}
		b_{n-1}&\cdots b_1 & b_0
		\end{bmatrix}x(t) + b_n y(t).
		\end{align*}
		
		\item Proportional-Integral-Derivative (PID) controller:
		With the transfer function given as the parallel form, as
		$$ C(z) = k_p + \frac{k_i T_s}{z-1} + \frac{k_d}{\frac{T_s}{N_d} + \frac{T_s}{z-1}},$$
		where $k_p$, $k_i$, and $k_d$ are the proportional, integral, and derivative gains, respectively, 	$T_s$ is the sampling time, and $N_d \in\N$ is the parameter for the derivative filter.
		It can be realized as
		\begin{align*}
		x(t+1) &= {\small\begin{bmatrix}
		2-N_d & N_d -1 \\
		1 & 0 
		\end{bmatrix}}x(t) + {\small\begin{bmatrix}
		1\\0
		\end{bmatrix}} y(t)\\
		u(t) &= \begin{bmatrix}
		b_1 & b_0
		\end{bmatrix} x(t) + b_2 y(t),
		\end{align*}
		where $b_1 = k_i T_s - k_d N_d^2/T_s$, $b_0 = k_i T_s N_d - k_i T_s + k_d N_d^2/T_s$, and $b_2 = k_p + k_d N_d/T_s$.
			\hfill$\square$
	\end{itemize}
\end{rem1}

In the remaining,
we consider the case of utilizing additively homomorphic encryption.
Implementation of \eqref{eq:controller_converted_mod} requires both multiplication and addition,
but
provided that the information of the matrices in \eqref{eq:controller_converted} is disclosed,
it can be encrypted, by exploiting the property H1--H3 of the cryptosystem only.
Indeed,
let the encrypted controller \eqref{eq:controller_encrypted} be replaced with
\begin{align}
\zz(t+1) &= F'\cdot \zz(t) + \ol G'\cdot \yy(t) + \ol P'\cdot \rr(t) + \ol \RRR\cdot\uu'(t)\notag\\
\uu(t) &= \ol H' \cdot \zz(t) + \ol J \cdot \yy(t) + \ol Q \cdot \rr(t),\label{eq:controller_proposed_additive}
\end{align}
where $\zz(0) = \Enc(\ol z(0)/\LLL\mod q)$,
and the operation $\cdot$ is based on the properties H3 and \eqref{eq:IntMatMult}.
Then, the following corollary states that
controller \eqref{eq:controller_proposed_additive} is also able to operate for an infinite time horizon, with the equivalent performance.
\begin{cor}
	If $\gamma(\LLL,\rrr_1,\rrr_2,\sss_1,\sss_2)\le \theta(\epsilon)$,
	then the controller \eqref{eq:controller_proposed_additive} with \eqref{eq:controller_encrypted_u} guarantees that $\|u(t)-u'(t)\|\le\epsilon$, $\forall t\in\N_0$.
\end{cor}

{\it Proof:}
Let $\tilde \zzz_e(t):=\Dec(\zz(t))\in\Z_q^{\nnn'}$
and
$\tilde \uuu_e(t):=\Dec(\uu(t))\in\Z_q^\mmm$. Then,
by H1 and \eqref{eq:IntMatMult},
they obey \eqref{eq:controller_encryption_decrypted}, with $\Delta_\Mult=0$.
Thus, Theorem~\ref{thm:main} completes the proof.
\hfill$\blacksquare$

And, in case the cryptosystem does not inject errors, i.e., $\Delta_\Enc=0$,
as the Paillier cryptosystem \cite{Paillier},
it directly follows that the performance of \eqref{eq:controller_proposed_additive} becomes the same with
\eqref{eq:controller_converted_integer}, where the factor $1/\LLL$
for the injected errors
is dispensable.
\begin{cor}
	Consider the controllers \eqref{eq:controller_proposed_additive} and \eqref{eq:controller_converted_integer} in their respective closed-loops, and
	suppose that
	$\Delta_\Enc=0$, $1/\LLL=1$, and $\beta(\rrr_1,\rrr_2,\sss_1,\sss_2 ) \le \theta(\epsilon)$.
	Then, they satisfy $ \ol u(t)=\Dec(\uu(t))\mod U$, for all $t\in\N_0$.
\end{cor}

\subsection{Parameter Design for Performance}
In this subsection,
we provide a guideline for choosing the parameters $\{\LLL,\rrr_1,\rrr_2,\sss_1,\sss_2\}$, which keeps the encrypted controllers from performance degradation.
Suppose
that the controller \eqref{eq:quantized_rev} over quantized numbers as well as a set $\{\rrr'_1,\rrr'_2,\sss'_1,\sss'_2\}$ of parameters
be given,
which satisfy the condition $\alpha'(\rrr_1',\rrr'_2,\sss'_1)\le \eta(\epsilon)$ defined in \eqref{eq:alpha}, in Proposition~\ref{prop:quantized_rev}.
Although
the condition $\gamma(\LLL,\rrr_1,\rrr_2,\sss_1,\sss_2)\le \theta(\epsilon)$ for guaranteeing the performance of \eqref{eq:controller_encrypted} can always be satisfied
with appropriate choice of parameters,
since $\gamma(0,0,0,0,0)=0$,
the choices of parameters should be prioritized, in practice.

Here, we suggest that the choice of $\rrr_1$ and $\rrr_2$ should be of priority.
This is because
increasing the parameters $1/\rrr_1$ and $1/\rrr_2$ of quantization may cost more than the others, as they may be determined from specification of the sensors and actuators.
Taking this into account,
we first
choose the parameters $1/\rrr_1$ and $1/\rrr_2$ with respect to \eqref{eq:controller_encrypted},
as small as possible;
we
choose $1/\rrr_1$ and $1/\rrr_2$
such that
\begin{align}
\frac{1}{\rrr_1}&>{\small\frac{\max\{1,\|\TTT'\|(1+\|\RRR\|)\}\cdot\max\{\|S\|,\| [J,Q]\|\}}{\max\{ \|[G,P] \|, \|[J,Q] \| \}}\cdot\frac{1}{\rrr_1'}},\notag\\
\frac{1}{\rrr_2}&>
	{\small\max\{1,\|\TTT'\|(1+\|\RRR\|)\}
\cdot\frac{1}{\rrr_2'}},\label{eq:gamma_prime}
\end{align}
where $S = [\TTT G - \RRR J, \TTT P - \RRR Q, \RRR]$,
so that from \eqref{eq:alpha}, \eqref{eq:theta}, \eqref{eq:e_z}, \eqref{eq:e_u},
\eqref{eq:beta},
and 
\eqref{eq:thm1_beta},
it guarantees that
\begin{equation}\label{eq:alpha0}
 \alpha' (\rrr'_1,\rrr'_2,0)\cdot \frac{\theta(\epsilon)}{\eta(\epsilon)}>\gamma (0,\rrr_1,\rrr_2,0,0).
\end{equation}

Then, the following proposition shows that the rest $\{\LLL,\sss_1,\sss_2\}$
can be chosen to
guarantee the performance of \eqref{eq:controller_encrypted}.

\begin{prop}
	Given $\{\rrr'_1,\rrr'_2,\sss'_1\}$ such that $\alpha'(\rrr'_1,\rrr'_2,\sss'_1)\le \eta(\epsilon)$,
	and $\{\rrr_1,\rrr_2\}$ satisfying \eqref{eq:gamma_prime},
	there exist
	$1/\LLL\in\N$, $1/\sss_1\in\N$, and $1/\sss_2\in\N$
	such that $\gamma(\LLL,\rrr_1,\rrr_2,\sss_1,\sss_2)\le \theta(\epsilon)$.
\end{prop}

{\it Proof:}
Define $f(l):= \gamma(l,\rrr_1,\rrr_2,l,l)-\alpha'(\rrr'_1,\rrr'_2,\sss'_1)\cdot(\theta(\epsilon)/\eta(\epsilon))$.
If $f(1)\le 0$, there is nothing to prove. Let $f(1)>0$.
From \eqref{eq:alpha0}, it is clear that $f(0)<0$.
By continuity,
there exists
$1/\LLL\in\N$
such that $f(\LLL)\le 0$.
It ends the proof.
\hfill$\blacksquare$

\subsection{Case of Single-Output Controllers}
Single-output controllers
can be easily converted to have the state matrix as integers.
Let $u(t)\in\R$, i.e., $\mmm=1$, and
let the observable subsystem \eqref{eq:controller_transformed_reduced} with \eqref{eq:controller_transformed_output} of the given controller be transformed into the observable canonical form, as
\begin{align*}
z(t+1) &= {\small\begin{bmatrix}
0 & \cdots & 0 \!&\! a_0\\
1 & \cdots & 0 \!&\! a_1\\
\vdots & \ddots\! &\! \vdots & \vdots\\
0 & \cdots & 1 \!&\! a_{\nnn'-1}
\end{bmatrix}}z(t)
+G'y(t) + P'r(t) \notag\\
u(t)&=\begin{bmatrix}
0&\cdots&0&1
\end{bmatrix}z(t)+Jy(t)+Qr(t),
\end{align*}
where
$a_i\in\R$, $i=0,\cdots,\nnn'-1$, $G'\in\R^{\nnn'\times\ppp}$, and $P'\in\R^{\nnn'\times\qqq}$.
Then,
with any vector $[k_1,\cdots,k_{\nnn'}]^\top\in\Z^{\nnn'}$,
we have $ R:= [a_0-k_0,\cdots,a_{\nnn'}-k_{\nnn'}]^\top$ so that
it can be re-written as
\begin{align}
\begin{split}\label{eq:single}
z(t+1) &= {\small\begin{bmatrix}
0 & \cdots & 0 \!&\! k_0\\
1 & \cdots & 0 \!&\! k_1\\
\vdots & \ddots\! &\! \vdots & \vdots\\
0 & \cdots & 1 \!&\! k_{\nnn'-1}
\end{bmatrix}}z(t)
+S \begin{bmatrix}
y(t)\\r(t)\\u(t)
\end{bmatrix} \\
u(t)&=\begin{bmatrix}
0&\cdots&0&1
\end{bmatrix}z(t)+Jy(t)+Qr(t),
\end{split}
\end{align}
with $S = [G'-RJ, P'-RQ, R]$,
where the state matrix is easily converted to integers.

\begin{rem}
	A benefit of the conversion \eqref{eq:single} is that the output matrix as well as the state matrix is converted to integers.
	Since we can
	have the scale factor $1/\sss_2$ for the output matrix	
	as $1/\sss_2=1$ for this case,
	it may reduce the modulus $q$,
	in practice, when it is determined by \eqref{eq:N} and \eqref{eq:input_integer_range}.
\end{rem}
\begin{rem}\label{rem:poleplace}
As analogous to the described conversion method for single-output controllers,
pole-placement techniques can also be considered for the conversion, in practice;
for example,
given the controller \eqref{eq:controller_given}
where $(F,H)$ is observable and $H\in\R^{1\times\nnn}$,
a matrix $R\in\R^{\nnn\times 1}$
can be first found such that
the coefficients of
the characteristic polynomial of the matrix $F-R H$ are all integers.
Then, considering the transfer function of the controller with respect to the form \eqref{eq:outward},
it can be realized as the form \eqref{eq:single}.
\end{rem}

\section{Simulation Results}\label{sec:simulation}
This section provides simulation results of the proposed scheme applied
to tracking control of three inertia system \cite{Ogata};
let the plant \eqref{eq:plant} be given as
\begin{align*}
\dot x_p(\ttt) &=
{\small\begin{bmatrix}
0\!&\!1\!&\!0\!&\!0\!&\!0\!&\!0\\
\frac{-{\sf k}}{\sf J}\!&\!\frac{-{\sf b}}{\sf J}\!&\!\frac{{\sf k}}{\sf J}\!&\!0\!&\!0\!&\!0\\
0\!&\!0\!&\!0\!&\!1\!&\!0\!&\!0\\
\frac{{\sf k}}{\sf J}\!&\!0\!&\!\frac{-2{\sf k}}{\sf J}\!&\!\frac{-{\sf b}}{\sf J}\!&\!\frac{{\sf k}}{\sf J}\!&\!0\\
0\!&\!0\!&\!0\!&\!0\!&\!0\!&\!1\\
0\!&\!0\!&\!\frac{{\sf k}}{\sf J}\!&\!0\!&\!\frac{-{\sf k}}{\sf J}\!&\!\frac{-{\sf b}}{\sf J}
\end{bmatrix}}x_p(\ttt) + {\small\begin{bmatrix}
0\\
\frac{1}{\sf J}\\
0\\
0\\
0\\
0
\end{bmatrix}}u_p(\ttt)\\
&=: A_p x_p(\ttt) + B_p u_p(\ttt)\\
y_p(\ttt) &= \begin{bmatrix}
0&0&0&0&1&0
\end{bmatrix}x_p(\ttt)=: C_p x_p(\ttt)
\end{align*}
where ${\sf J}= 0.01\, \mathrm{kg}\!\cdot\! \mathrm{m}^2$, ${\sf b} = 0.007\, \mathrm{N}\!\cdot\!\mathrm{m}/(\mathrm{rad}/\mathrm{s})$,
and ${\sf k}= 1.37\,\mathrm{N}\!\cdot\!\mathrm{m}/\mathrm{rad}$,
and the system \eqref{eq:controller_given} be designed as an observer-based discrete-time feedback controller with $T_s = 50\,\mathrm{ms}$, as
\begin{align}\label{eq:sim}
\hat x_p(t+1) &= (A_d +B_dK-LC_p)\hat x_p(t)+B_d K_I \xi(t) + Ly(t)\notag\\
\xi(t+1) &= \xi(t)  - C_p \hat x_p(t)+ r(t),\quad x(t) = [\hat x_p(t)^\top , \xi(t)]^\top\notag\\
u(t) &= K \hat x_p(t) + K_I \xi (t)=Hx(t),
\end{align}
where
$A_d = \exp({A_pT_s})$ and $B_d =\int_0^{T_s}\exp({A_p\tau})d\tau\cdot B_p $,
and
\begin{align*}
	\begin{split}
K&=-\begin{bmatrix}
	2.32,~ 0.25,~ -2.47,~ 0.04,~ 1.70,~ 0.12
\end{bmatrix}, ~K_I=0.1,\\
L &= \begin{bmatrix}
0.47,~
4.07,~
-0.03,~
-6.80,~
1.39,~
6.21
\end{bmatrix}^\top,
	\end{split} 
\end{align*}
are the proportional and integral feedback gains, and
the injection gain for the observer,
respectively,
so that the estimate $\hat x_p(t)$ converges to the plant state $x_p(t\cdot T_s)$, and the output $y_p(\ttt)$ of the plant follows the reference $r(t)\equiv1$.

The state matrix
$F\in\R^{7\times 7}$ of \eqref{eq:sim} is converted with the method of Remark~\ref{rem:poleplace};
the matrix $R$ is found such that $$\det(zI_{7}-(F-R H))=z^7 -3z^6 +3z^5 -3z^4 +z^3 -1, $$
so that \eqref{eq:sim}
is converted to the form
\eqref{eq:single}, with
$${\small\begin{bmatrix}
k_0\\k_1\\k_2\\k_3\\k_4\\k_5\\k_6
\end{bmatrix}}={\small\begin{bmatrix}
1\\0\\0\\-1\\3\\-3\\3
\end{bmatrix}},~
S
= {\small\begin{bmatrix}
-2.5357   \!\!&\!\!    0.0108	\!\!&\!\! -0.9931\\
16.0183   \!\!&\!\!   -0.0737	\!\!&\!\!-0.0794\\
-43.0087   \!\!&\!\!    0.2305	\!\!&\!\! 0.4673\\
62.9373   \!\!&\!\!   -0.4243	\!\!&\!\!-0.3812\\
-53.2140  \!\!&\!\!     0.4849	\!\!&\!\!-0.3892\\
24.8186  \!\!&\!\!    -0.3254	\!\!&\!\!-0.4425\\
-5.0182   \!\!&\!\!   0.1000	\!\!&\!\!-0.1886 
\end{bmatrix}}. $$
	\begin{figure}[tp]
		\centering
		\includegraphics[width=0.85\columnwidth]{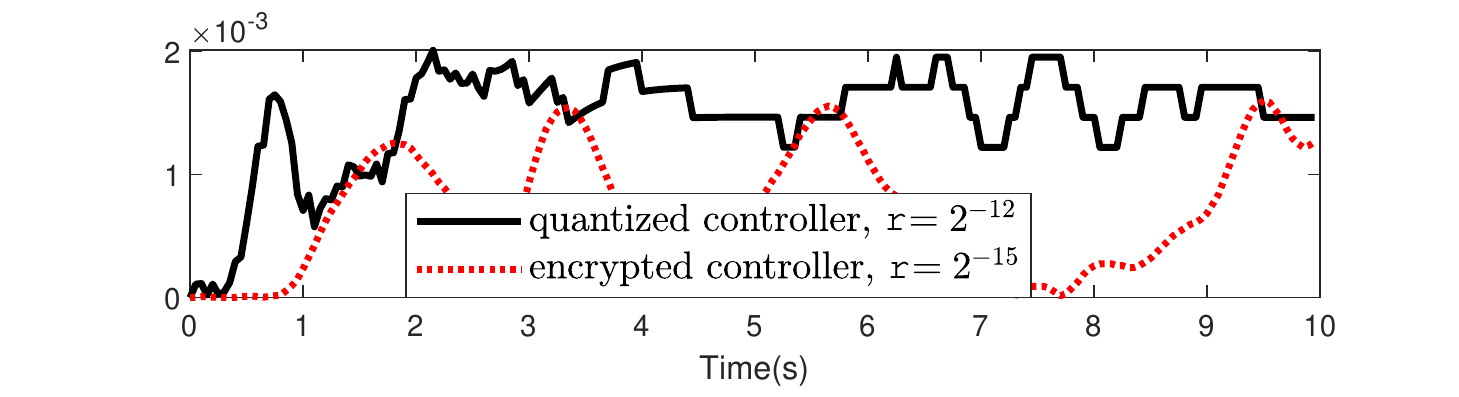}	
		\caption{Performance error $\|y(t) - y'(t)\|$ of the quantized controller with
			$\rrr_1=\rrr_2=\mathrm{r}=2^{-12}$
			and $\sss_1=\sss_2=2^{-12}$
			(black line),
			and of the encrypted controller with
			$\rrr_1=\rrr_2=\mathrm{r}=2^{-15}$
			and 
			$(\LLL,\sss_1,\sss_2)=(2^{-11},2^{-19},1)$
			(red dotted-line).
		}
		\label{fig1}
	\end{figure}
\begin{figure}[tp]
	\centering
	\includegraphics[width=0.85\columnwidth]{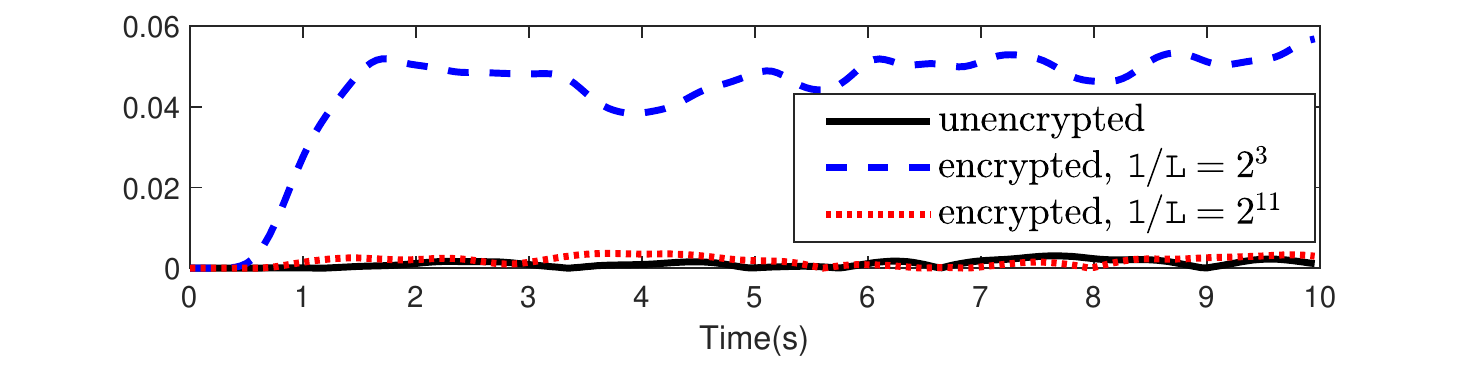}	
	\caption{Performance error $\|y(t) - y'(t)\|$ of the converted controller
		with
		$(\rrr_1,\rrr_2,\sss_1,\sss_2)=(2^{-15},2^{-15},2^{-19},1)$
		(black line),
	and that of
	the encrypted controllers
		with
		the same parameters,
		with
		$1/\LLL=2^3$
		(blue dashed-line) and
		with
		$1/\LLL=2^{11}$
		(red dotted-line), respectively.
	}
	\label{fig2}
\end{figure}

{\color{black}	
For the operation, the described cryptosystem \cite{GSW-LWE} is used
with the parameters $(q,\sigma,n)=(2^{48},1,249)$, where the total average time for the encryption, control operation, and decryption was within the sampling period $T_s = 50\mathrm{ms}$.
}

\begin{figure}
	\centering
	\begin{subfigmatrix}{4}
		~~~~\subfigure[]{\includegraphics[width=.16\textwidth]{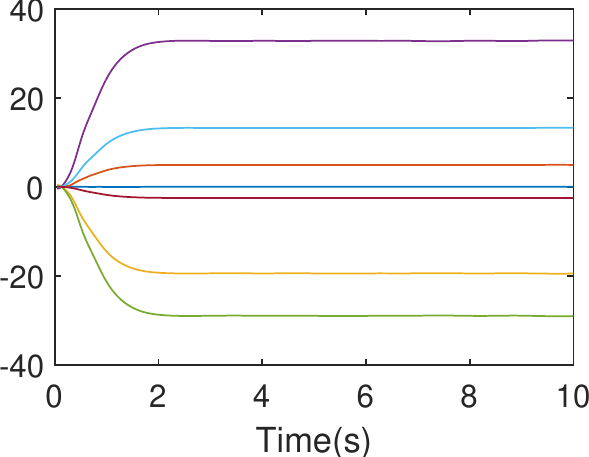}\label{fig:unencrypted}}\!\!\!\!\!\!\!\!\!\!\!\!\!\!\!\!\!\!\!\!\!
		\subfigure[]{\includegraphics[width=.16\textwidth,keepaspectratio]{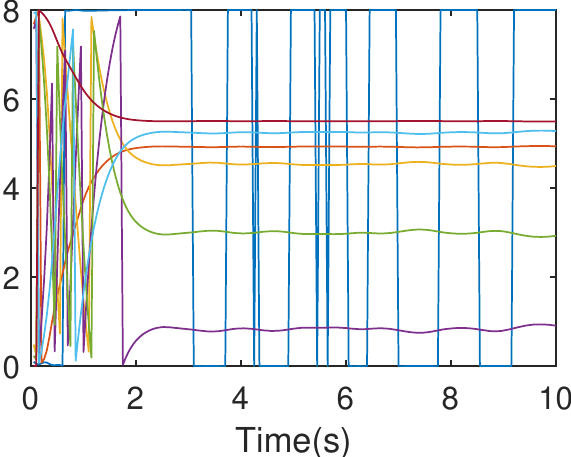}\label{fig:encrypted}}~~~~~\\
		\centering
		\subfigure[]{\includegraphics[width=.16\textwidth,keepaspectratio]{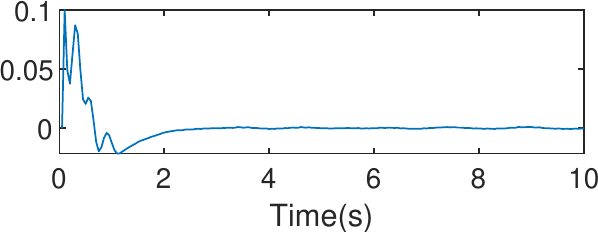}\label{fig:unencrypted_u}}~~~~~
		\subfigure[]{\includegraphics[width=.16\textwidth,keepaspectratio]{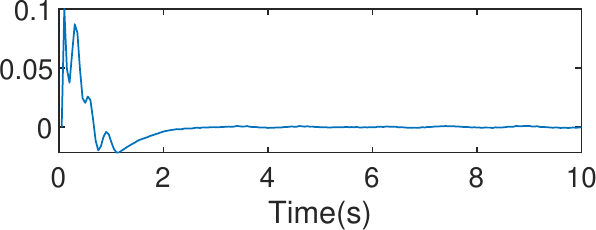}\label{fig:encrypted_u}}~~~~~~
		\end{subfigmatrix}
	\caption{(a) Plot of the state $\LLL\rrr_1\sss_1\cdot \tilde z(t)$ of the converted controller \eqref{eq:controller_converted_scaled_integer}.
(b)	Plot of the state $\LLL\rrr_1\sss_1\cdot\Dec(\zz(t))$ of the encrypted controller \eqref{eq:controller_encrypted}.
(c) Output $u(t)$ of the converted controller \eqref{eq:controller_converted_scaled_integer}.
(d) Output $u(t)$ of the encrypted controller \eqref{eq:controller_encrypted}.
}
\label{fig:cut}
\end{figure}

Fig.~\ref{fig1}, \ref{fig2}, and \ref{fig:cut} show the simulation results of the proposed encrypted controller,
which operates without reset or decryption of the state.
Compared with the method re-encrypting the state,
the amount for the re-encryption is reduced by a seventh, since $\nnn=7$ and $\mmm=1$.
Compared with the quantized controller,
Fig.~\ref{fig1} shows that
the performance of the proposed encrypted controller can be preserved with the choice of parameters and can perform unlimited operation, as proposed in Theorem~\ref{thm:main}.
Fig.~\ref{fig2} supports the statement of Remark~\ref{rem:L};
increasing the parameter $1/\LLL$ only, it suppresses the effect of injected errors and obtains the same level of performance with the unencrypted model.
And, Fig.~\ref{fig:cut} demonstrates the effectiveness of the proposed criterion for choosing the modulus $q$ in a less conservative way.
Despite that the higher bits of the message in the encrypted state are cut off by modulo operation, it keeps its output performance, as the same.

\section{Conclusion}\label{sec:conclu}
In this paper,
we have proposed that linear dynamic controllers can be encrypted to operate for an infinite time horizon with the same level of performance,
when the transmission of the encrypted plant input to the controller is admitted.
The proposed method simply converts the state matrix of the given controller by pole-placement design,
so that it fixes the scale of the control variables
and eliminates the necessity of the reset of the state.
To conceal the control parameters as well as the signals,
the use of LWE-based cryptosystem has been considered,
in which it has been seen that the effect of injecting errors to the messages is controlled by closed-loop stability.

{

	\renewcommand{\baselinestretch}{0.9}

}

\begin{IEEEbiography}[{\includegraphics[width=1in,height=1.25in,clip]{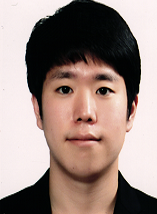}}]{Junsoo Kim}
	received his B.S. degree in electrical engineering and mathematical sciences in 2014, and M.S. and Ph.D. degrees in electrical engineering in 2020, from Seoul National University.
	From 2020 to 2021, he held the post-doc position at Automation and Systems Research Institute, Korea.
	And, he is currently a postdoctoral researcher at KTH Royal Institute of Technology, Sweden.
	His research interests include security problems in networked control systems and encrypted control systems.
\end{IEEEbiography}
\begin{IEEEbiography}
	[{\includegraphics[width=1in,height=1.25in,clip]{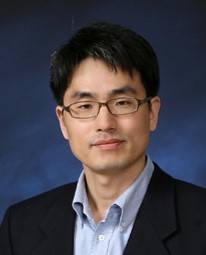}}]
	{Hyungbo Shim}
	received the B.S., M.S., and Ph.D. degrees from Seoul National University, Korea, and held the post-doc position at University of California, Santa Barbara till 2001. He joined Hanyang University, Seoul, in 2002. Since 2003, he has been with Seoul National University, Korea. He served as associate editor for Automatica, IEEE Trans. on Automatic Control, Int. Journal of Robust and Nonlinear Control, and European Journal of Control, and as editor for Int. Journal of Control, Automation, and Systems.
\end{IEEEbiography}
\begin{IEEEbiography}[{\includegraphics[width=1in,height=1.25in,clip]{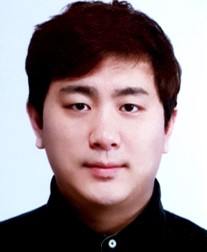}}]
{Kyoohyung Han}
received the Ph.D. degree in mathematical sciences from Seoul National University, Seoul, South Korea, in 2019. After that, he was a Post-Doctoral researcher in the Research Institute of Basic Sciences at Seoul National University. He is currently a researcher at Samsung SDS, South Korea. His research interests include cryptographic primitives for secure computation and privacy enhancing techniques.
\end{IEEEbiography}

\begin{thebibliography}{99}
	
	\bibitem{Kogiso15CDC}
	K.~Kogiso and T.~Fujita, ``Cyber-security enhancement of networked control systems using homomorphic encryption," in {\it Proc.~54th IEEE Conf.~Decision and Control}, 2015, pp.~6836--6843.	
	
	\bibitem{Kim16NECSYS}
	J.~Kim, C.~Lee, H.~Shim, J.\,H.~Cheon, A.~Kim, M.~Kim, and Y.~Song, ``Encrypting controller using fully homomorphic encryption for security of cyber-physical systems,'' {\it IFAC-PapersOnline}, vol.~49, iss.~22, pp.~175--180, 2016.
	
	\bibitem{Farokhi17CEP}
	F.~Farokhi, I.~Shames, and N.~Batterham, ``Secure and private control using semi-homomorphic encryption,'' {\it Control Engineering Practice}, vol.~67, pp.~13--20, 2017.
	
	\bibitem{Shoukry16CDC}
	A.\,B.~Alexandru, K.~Gatsis, Y.~Shoukry, S.\,A.~Seshia, P.~Tabuada, and G.\,J.~Pappas, ``Cloud-based quadratic optimization with partially homomorphic encryption,'' {\it IEEE Trans.~on Automatic Control}, vol.~66, no.~5, pp.~2357--2364, 2021.
	
	\bibitem{Hadjicostis18CDC}
	C.\,N.~Hadjicostis and A.\,D.~Dom{\'i}nguez-Garc{\'i}a,
	``Privacy-preserving distributed averaging via homomorphically encrypted ratio consensus,''
	{\it IEEE Trans.~on Automatic Control}, vol.~65, no.~9, pp.~3887--3894, 2020.
	
	
	
	\bibitem{Darup19CSL}
	M.~Schulze~Darup, A.~Redder, and D.\,E.~Quevedo, ``Encrypted cooperative control based on structured feedback,'' {\it IEEE Control~Syst.~Lett.}, vol.~3, iss.~1, pp.~37--42, 2019.
	
	\bibitem{Alexandru18}
	A.\,B.~Alexandru, M.~Morari, G.\,J.~Pappas,
	``Cloud-based MPC with encrypted data," in {\it Proc.~57th IEEE Conf.~Decision and Control}, 2018, pp.~5014--5019.
	
	\bibitem{Murguia18Arxiv}
	C.~Murguia, F.~Farokhi, and I.~Shames, ``Secure and private implementation of dynamic controllers using semi-homomorphic encryption,"
	{\it IEEE Trans.~on Automatic Control}, vol.~65, no.~9, pp.~3950--3957, 2020.
	
	\bibitem{Gentry09}
	C.~Gentry,
	``Fully homomorphic encryption using ideal lattices,''
	in {\it Proc. STOC}, vol.~9, 2009, pp.~169--178.
	
	\bibitem{Cheon18}
	J.\,H.~Cheon, K.~Han, H.~Kim, J.~Kim, and H.~Shim,
	``Need for controllers having integer coefficients in homomorphically encrypted dynamic system," in {\it Proc.~57th IEEE Conf.~Decision and Control}, 2018, pp.~5020--5025.
	
	
	\bibitem{Paillier}
	P.~Paillier, ``Public-key cryptosystems based on composite degree residuosity classes,'' in {\it Proc.~17th Int.~Conf.~Theory~Applicat.~Crypto.~Tech.}, 1999, pp.~223--238.
	
	\bibitem{Reg05}
	O.~Regev,
	``On lattices, learning with errors, random linear codes, and cryptography,'' {\it Journal of the ACM}, vol.~56, no.~6, pp.~34, 2009.
	
	
	
	\bibitem{GSW13}
	G.~Gentry, A.~Sahai, and B.~Waters,
	``Homomorphic encryption from learning with errors:
	conceptually-simpler, asymptotically-faster, attribute-based,''
	in
	{\it Advances in Cryptology--CRYPTO}, Springer, Berlin, Heidelberg, 2013, pp.~75--92.
	
	\bibitem{GSW-LWE}
	I.~Chillotti, N.~Gama, M.~Georgieva, M.~Izabach{\`e}ne,
	``Faster fully homomorphic encryption: bootstrapping in less than 0.1 seconds,''
	in
	{\it Advances in Cryptology--ASIACRYPT 2016}, Springer, Berlin, Heidelberg, 2016, pp.~3--33.
	
	\bibitem{quantum}
	L.~Chen, S.~Jordan, Y.\,K.~Liu, D.~Moody, R.~Peralta, R.~Perlner, D.~Smith-Tone,
	{\it Report on post-quantum cryptography},
	US Department of Commerce, National Institute of Standards and Technology, 2016.
	
	\bibitem{Tran19}
	J.~Tran, F.~Farokhi, M.~Cantoni, I.~Shames,
	``Implementing homomorphic encryption based secure feedback control for physical systems,''
	 {\it Control Engineering Practice}, vol.~97, 2020, 104350.	
	
	\bibitem{Kim19}
	J.~Kim, H.~Shim, and K.~Han,
	``Comprehensive introduction to fully homomorphic encryption for dynamic feedback controller via LWE-based cryptosystem,''
	in
	{\it Privacy in Dynamical Systems},
	pp.~209--230,
	Springer, Singapore, 2020.	ArXiv:1904.08025 [cs.SY].
	
	\bibitem{Cheon17ASIACRYPT}
	J.\,H.~Cheon, A.~Kim, M.~Kim, and Y.~Song,
	``Homomorphic encryption for arithmetic of approximate numbers,'' in {\it Proc.~Int.~Conf.~Theory.~Applicat.~Crypto.~Inform.~Security}, 2017, pp.~409--437.
	
	
	
	
	\bibitem{Kim19Arxiv}
	J.~Kim, H.~Shim, and K.~Han,
	``Dynamic controller that operates over homomorphically encrypted data for infinite time horizon,'' arXiv:1912.07362v1 [eess.SY], 2019.
	
	
	\bibitem{Kim20CDC}
	J.~Kim, H.~Shim, and K.~Han,
	``Design procedure for dynamic controllers based on LWE-based homomorphic encryption to operate for infinite time horizon,''
	in {\it Proc.~59th IEEE Conf.~Decision and Control}, 2020, pp.~5463--5468.
		
	\bibitem{Ogata}
	K.~Ogata, {\it Discrete-Time Control Systems}, 2nd ed. Englewood Cliffs, NJ,
	USA: Prentice Hall, 1995.
	
	
	
	

\end{thebibliography}
\end{document}